\documentclass[
 amsmath,amssymb,
 aps, prl, twocolumn, superscriptaddress
]{revtex4-2}

\usepackage{amsmath,amssymb}
\usepackage{graphicx}
\usepackage{xcolor}
\usepackage{bm}
\usepackage{csquotes}

\newcommand{\mb}[1]{\mathbf{#1}}

\providecommand{\abs}[1]{\left\lvert #1 \right\rvert}

\def\[{\left[}
\def\]{\right]}
\def\({\left(}
\def\){\right)}

\def\figurename{Fig.}
\makeatletter
\let\@orig@make@capt@title\@make@capt@title
\def\@make@capt@title#1#2{\@orig@make@capt@title{{\bf #1}}{#2}}
\makeatother

\usepackage{hyperref}
\usepackage{bookmark}

\definecolor{tab_blue}{HTML}{1F77B4}
\definecolor{tab_red}{HTML}{D62728}

\hypersetup{%
  breaklinks=true,
  colorlinks=true,
  linkcolor=tab_blue,
  filecolor=tab_blue,
  urlcolor=tab_blue,
  citecolor=tab_blue,
}

\begin{document}
\title{Learning dynamical behaviors in physical systems}

\newcommand{\jfi}
{\affiliation{James Franck Institute, The University of Chicago, Chicago, IL 60637}}

\newcommand{\phy}
{\affiliation{Department of Physics, The University of Chicago, Chicago, IL 60637}}

\newcommand{\kadanoff}
{\affiliation{Kadanoff Center for Theoretical Physics, The University of Chicago, Chicago, Illinois 60637, USA}}

\newcommand{\gulliver}
{\affiliation{Gulliver, ESPCI Paris, Université PSL, CNRS, Paris, France}}

\newcommand{\chem}
{\affiliation{Department of Chemistry, University of Chicago, Chicago, Illinois 60637, USA}}

\author{Rituparno Mandal}%
\thanks{These authors contributed equally to this work.}
\jfi

\author{Rosalind Huang}%
\thanks{These authors contributed equally to this work.}
\jfi
\phy

\author{Michel Fruchart}%
\gulliver

\author{Pepijn G. Moerman}%
\affiliation{Chemical Engineering and Chemistry, Eindhoven University of Technology, Eindhoven 5612 AP, Netherlands}

\author{Suriyanarayanan Vaikuntanathan}%
\jfi
\chem

\author{Arvind Murugan}
\email{amurugan@uchicago.edu}%
\phy

\author{Vincenzo Vitelli}%
\email{vitelli@uchicago.edu}
\jfi
\phy
\kadanoff

\begin{abstract}
Physical learning is an emerging paradigm in science and engineering whereby (meta)materials
acquire desired macroscopic behaviors by exposure to examples.
So far, 
it has been applied to static properties such as elastic moduli and self-assembled structures encoded in minima of an energy landscape.
Here, we extend this paradigm to dynamic functionalities, such as motion and shape change,  
that are instead encoded in limit cycles or pathways of a dynamical system.
We identify the two ingredients needed to learn time-dependent behaviors irrespective of experimental platforms: (i) learning rules with time delays and (ii) exposure to examples that break time-reversal symmetry during training. After providing a hands-on demonstration of these requirements using programmable LEGO toys, we turn to realistic particle-based simulations where the training rules
are not programmed on a computer. 
Instead, we elucidate how they emerge from physico-chemical processes involving the causal propagation of fields, like in recent experiments on moving oil droplets with chemotactic signalling. 
Our trainable particles can 
self-assemble into structures that move or change shape on demand, either by retrieving the dynamic behavior previously seen during training, or by learning on the fly.
This rich phenomenology is captured by a modified Hopfield spin model amenable to analytical treatment.
The principles illustrated here provide a step towards von Neumann's dream of engineering synthetic living systems that adapt to the environment.
\end{abstract}

\maketitle

Biological systems adapt to their environment by changing their behavior in response to past events: once bitten, twice shy.  In the brain, for instance, neurons wire together (i.e. increase their interactions) if they fire together.
But the ability to learn is not limited to sentient or animate beings: it can emerge in natural physical processes in which interactions increase between entities that happen to be correlated, a paradigm known as Hebbian learning~\cite{hebb_organization_2002,hebb_organization_2002}. 
Inanimate systems too can evolve their microscopic interactions to effectively learn desired behaviors after experiencing examples of that behavior, a phenomenon often referred to as physical learning~\cite{stern_learning_2023,Keim2019,Scheibner2022}.
So far, 
physical learning has been applied to static behaviors, such as elastic responses or self-assembled shapes and particles configurations
characteristic of equilibrium systems~\cite{murugan_multifarious_2015,Falk2023b,Pashine2019,Stern2022,McMullen2022,Glotzer2004,Lee2022,Djellouli2024,McMullen2022,Evans2024,Wang2023,Aguilar2016,Glotzer2004,King2023,Niu2019}. Such static targets are encoded in fixed points, i.e. local minima of an energy landscape. 

\begin{figure*}[t]
{\centering
\includegraphics[width=\textwidth]{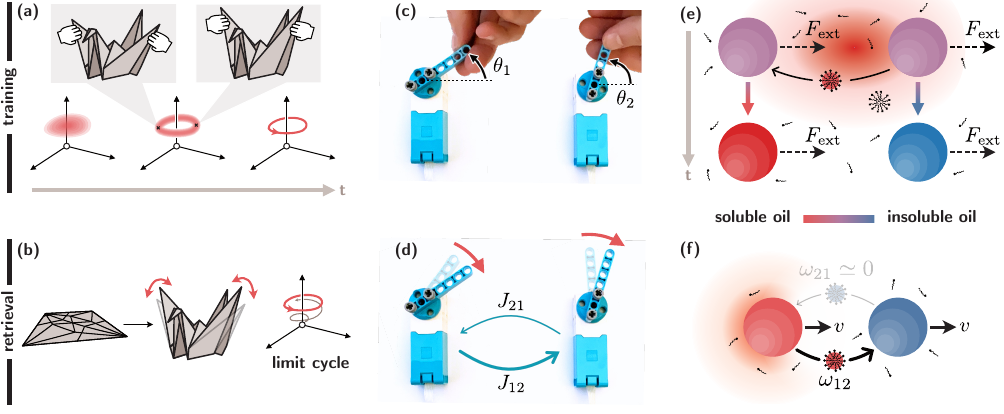}}
\caption{\label{fig_intro}
\textbf{Learning dynamic states.}
Panels a-b show the general strategy: during a training period (panel a), a time-dependent state is imposed on the system. 
This is pictured by an origami bird whose wings are periodically moved by an external agent.
After training (panel b), a retrieval period during which the physical system evolves under the learned dynamics should lead to a dynamic state, such as a limit cycle, that matches the training as best as possible.
(c-d) Demonstration of the training and retrieval procedure using a programmable LEGO toy (see also Movie 1). The angular positions $\theta_1(t)$ and $\theta_2(t)$ of two motors are imposed by hand during a training phase (panel c), during which couplings $J_{12}$ and $J_{21}$ between the motors are learned. Dynamics during retrieval, with fixed couplings $J_{ij}$,  can produce fixed points as well as dynamic states where the angles constantly change (panel d).
(e-f) We sketch how this learning and retrieval procedure can be implemented in a model physical system composed of moving oil droplets that interact through Marangoni flows~\cite{Meredith2020,Moran2017,Michelin2023,Nguyen2002,Maass2016,Michelin2013,Izri2014,Schmitt2013,Soto2014,Moerman2017}. 
The droplets are composed of a mixture of an oil that can easily be solubilized in water by micelles (in red) and a less soluble oil (in blue).
During training (panel e), two droplets with identical compositions (in purple) are driven by an external force (e.g. optical tweezers) at high enough Péclet number, leading to a comet-like diffusion cloud of oil behind the droplets. 
(The Péclet number is $\text{Pe} = a v_{\text{ext}}/D$ in which $a$ is the separation between the particles, $v_{\text{ext}}$ the velocity during training, and $D$ the diffusion coefficient of the solubilized oil.)
As a consequence, the composition of both droplets changes (middle row), with the front droplet having more insoluble oil and the back droplet more soluble oil.
During retrieval (panel f), we start with two droplets with different compositions. These can have non-reciprocal interactions leading to a run-and-chase dynamics originating from Marangoni effects~\cite{Meredith2020}.
}
\end{figure*}

In this article, we ask: how can a physical system learn time-dependent functionalities like pathways, trajectories, or dynamic states such as limit cycles?
The essence of our approach is conceptualized schematically in Fig.~\ref{fig_intro}: an external operator first imposes a dynamic state, that breaks time-reversal symmetry, on the system during a period called training (e.g. moves the wings of the origami bird in Fig.~\ref{fig_intro}a). 
Next, during a period that we call retrieval, the desired time-dependent state (e.g. wings flapping) can be recovered as the system evolves according to asymmetric microscopic interactions learned during training (Fig.~\ref{fig_intro}b).

Our foundational step is to identify the basic ingredients needed to learn dynamical states irrespective of experimental platforms. We begin with a hands-on demonstration of these requirements using the programmable LEGO toy shown in Fig.~\ref{fig_intro}c-d and Movie 1. 
The toy has two angular degrees of freedom $\theta_1$ and $\theta_2$ that can be controlled by hand very much like the angles of the origami-bird wings. These angles follow a dynamics enabled by sensors and actuators where each angle tends to align or antialign with the other depending on the tunable couplings $J_{12}$ and $J_{21}$ (here $J_{12}$ represents the effect of $\theta_2$ on $\theta_1$). 
During the training phase, the operator changes the angles by hand over time.
The time series of the angles are in turn mapped into distinct values of the couplings $J_{ij}$ by a learning rule programmed on a computer. 
The simplest Hebbian learning rule, in which $J_{i j}$ increases when $\theta_i$ and $\theta_j$ are simultaneously aligned, 
precludes dynamic states because only symmetric interactions ($J_{12} = J_{21}$) are learned. In order to retrieve dynamic states (Fig.~\ref{fig_intro}d), asymmetric couplings ($J_{12} \neq J_{21}$) are generically needed~\cite{Parisi1986,Crisanti1987,Sompolinsky1986,Kanter1987,Kleinfeld1986,Herron2023,Dehaene1987,Buhmann1987,Kleinfeld1988,Hertz1987,Hertz1986,Ivlev2015,Fruchart2021,Saha2020,You2020,Yifat2018,Drescher2009,Osat2022,Baek2018,banerjee2022,Dinelli2023,mandal2022,Avni2023}. 

Asymmetric couplings can be learned if a time delay~$\tau$ is added in the learning rule (Movie 1). For instance, we can impose that $J_{i j}$ tends to increase when $\theta_i(t)$ is correlated with $\theta_j(t-\tau)$. This creates the desired asymmetry between $i$ and $j$.
But it is not enough: the system must be trained using examples that break time-reversal symmetry, in order for the asymmetry not to be washed out over time (see Movie 1). 
Learning rules with signal-propagation delays naturally arise from causality and are reminiscent of spike-timing-dependent plasticity (STDP) in neuroscience~\cite{Abbott2000,Dan2004,caporale_spike_2008, shouval_spike_2010, song_competitive_2000-1,Fiete2010}: the influence of neuron $j$ on neuron $i$ increases if $j$ fires just before $i$ fires (but not if $j$ fires just after $i$).

\textit{Learning time-dependent states in physical systems.}---In the example of Fig.~\ref{fig_intro}c-d, a computer is needed to program the dynamics of the system during both training and retrieval.
We now demonstrate our approach in situations where the learning rule capable of learning non-reciprocal interactions is not programmed on a computer, but instead naturally emerge from basic physico-chemical processes.
Figure~\ref{fig_intro}e-f shows an example of trainable interaction between oil droplets~\cite{Meredith2020,Moran2017,Michelin2023,Maass2016,Michelin2013,Izri2014,Schmitt2013,Soto2014,Moerman2017,JambonPuillet2024}.
The droplets are composed of a mixture of two chemicals: the red oil can easily be solubilized in water, while the blue one is less soluble. 
It has been observed experimentally that this asymmetry in their composition can lead to a motion of the droplets in which the red droplet chases the blue one~\cite{Meredith2020}. 

The aforementioned process is what we need in our proposal of a concrete
training protocol for learning time-dependent states in droplets experiments. In the training phase, forces are applied on two droplets with identical initial composition (purple in panel e). 
When the droplets move fast enough (i.e. at large enough Péclet number, see caption and next section), the asymmetric diffusion of the red miscible oil due to the motion of the droplets (see the comet-like diffusion cloud in top part of panel e) leads to a change in their composition (bottom part of panel e).
Crucially, the red oil of the front droplet can be captured by the back droplet, but not vice-versa.
As a consequence, the droplets tend to reproduce the motion seen in training during the retrieval phase where they are left to freely evolve (panel f). 
We refer the reader to Methods for more detailed experimental considerations.

The key feature of the example of Fig.~\ref{fig_intro}e-f is the appearance of time-delayed interactions arising from the combination of the particles' motion during training with the causal propagation of signals through a field (e.g., oil concentration in Fig.~\ref{fig_intro}). 
Beyond this specific example, all physical interactions are subject to retarded potentials due to causality, for example asymmetric forces between acoustically levitated particles 
or hydrodynamic interactions mediated by the wakes of objects moving in a fluid~\cite{banerjee2022}.
We now consider a simplified particle-based model that captures this ubiquitous feature in a minimal mathematical setting amenable to analytic treatment. 

\begin{figure*}[t]
\centering
\hspace{-1em}\includegraphics[width = 1.72\columnwidth]{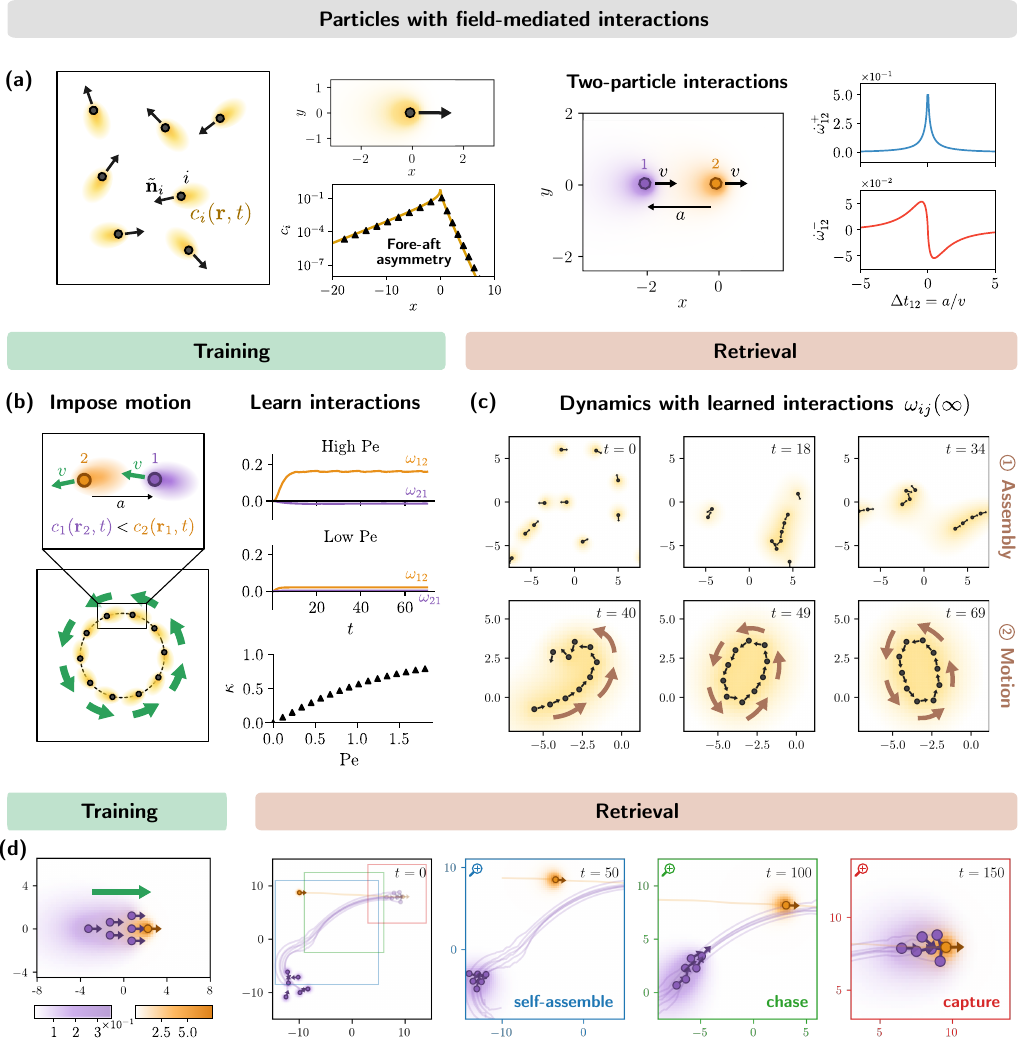}
\caption{\textbf{Learning dynamic states in a particle system.} 
(a) Proposed physical system to learn dynamic steady states. A collection of active, chemotactic particles that release chemical concentration $c_i(\mb{r},t)$. The chemical concentration field around a particle breaks fore-aft symmetry, with exponential decay having different decay length scale in front vs. behind the moving particle, obtained both from simulation and analytical solution (detailed in Methods). (b) Training protocols involve constrained motion of the particles while chemotactic affinities evolve. Particles learn chemotactic affinities according to concentrations they sense. The learned interactions can become symmetric (for $\text{Pe}=0.18$) or asymmetric (for $\text{Pe}=1.88$) depending on parameters such as diffusivity $D$ or velocity $v$ etc. The degree of asymmetry, defined as $\kappa=\frac{|\bar{\omega}_{12}-\bar{\omega}_{21}|}{|\bar{\omega}_{12}+\bar{\omega}_{21}|}$increases monotonically with $\text{Pe}$ (the overbar denotes an average over time and over all nearest neighbour pairs). (c) Using the learned affinities, the particles assemble and move. Snapshots from a simulation with $10$ particles and $\tau_c =1.0$, $\ell_c=0.5$, $\lambda=0.1$, $D=0.5$, $v=0.5$. (d) Training for a functional behavior where particles (violet) learn to assemble into a triangular formation and collectively chase a different target particle (orange). The training was performed for $v=1.5$, $l_c=0.5$, $\lambda=0.5$, $D=1.0$ for all the particles, where we used for the chasing ones $\tau_c=1.0$ and for the target $\tau_c=0.025$. During the retrieval we first the see formation of a triangular assembly which subsequently chases and captures the target.
}
\label{fig2}
\end{figure*}

\textit{Particles with field-mediated interactions.}\textemdash 
Consider particles at positions $\mb{r}_i(t)$ (Fig.~\ref{fig2}a) that are sources for distinct fields (e.g. chemicals such as different types of oil with distinct solubilities as in Fig.~\ref{fig_intro}e-f) with concentration $c_i(\bm{r},t)$ (shown in yellow in Fig.~\ref{fig2}a). 
These concentration fields evolve according to
\begin{align}
    \partial_t c_i(\mathbf{r},t) = \tau_{\text{c}}^{-1} \Theta(|\mathbf{r}-\mathbf{r}_i(t)|-\ell_{\text{c}}) -\lambda c_i + D \nabla^2 c_i.
    \label{dot_c}
\end{align}
The first term describes the emission of $c_i(\bm{r},t)$ with rate $\tau_{\text{c}}^{-1}$ in a circle of radius $\ell_{\text{c}}$ around particle $i$ ($\Theta$ is the Heaviside function). The second and third terms describe respectively decay at rate $\lambda$ and diffusion with diffusion constant $D$.

The positions of the particles follow the overdamped dynamics 
\begin{align}
    \frac{d\mathbf{r}_i}{dt} &= v_0 \Tilde{\mb{n}}_i - \sum_{j\neq i}\nabla_i V_{ij} +{\boldsymbol{\xi}}(t) \label{dot_r}.
\end{align}
with steric repulsion $V_{ij}$,  translational noise ${\boldsymbol{\xi}}(t)$, and self-propulsion with speed $v_0$ and a direction $\Tilde{\mb{n}}_i\propto \sum_{j\neq i} \omega_{ij} \nabla c_j(\mathbf{r}_i,t)$
set by the (generalized) mobilities $\omega_{ij}$, that describe the tendency of particle $i$ to move along gradients of $c_j$. The $\omega_{ij}$ could be, for example, chemotactic or diffusiophoretic mobilities whose values are fixed at the end of the training phase (see Methods for detailed modelling, e.g. the influence of angular inertia).

Inspired by the experimental proposal in Fig.~\ref{fig_intro}e-f, we formulate the following training rule that captures mathematically how the $\omega_{ij}$ evolve as a function of the trajectories $\bm{r}_i(t)$ that the operator imposes by hand
during training (overruling Eq.~\eqref{dot_r}):
\begin{align}
    \frac{d{\omega}_{ij}}{d t} &= \alpha\big[c_j(\mathbf{r}_i(t),t) - c_0\big] -\frac{\omega_{ij}}{\tau_0}. \label{dot_om}
\end{align}
Here $\alpha$ is the learning rate and $\tau_0$ is a relaxation time.
In a nutshell, this learning rule increases the mobility $\omega_{i j}$ when particle $i$ detects a greater-than-average value of $c_j$. 
In addition to the mechanism sketched in Fig.~\ref{fig_intro}, such strengthening of interactions based on spatial co-localization of chemical species or other constituents can be achieved experimentally by exploiting a family of proximity-based ligation, polymerization or other chemical mechanisms~\cite{stern_learning_2023,Fredriksson2002,Schaus2017,Moerman2023,Weinstein2019,Owen2023,Falk2023,Evans2024,murugan_multifarious_2015}.

\begin{figure*}[]
\centering
\hspace{-2em}\includegraphics[width = 1.75\columnwidth]{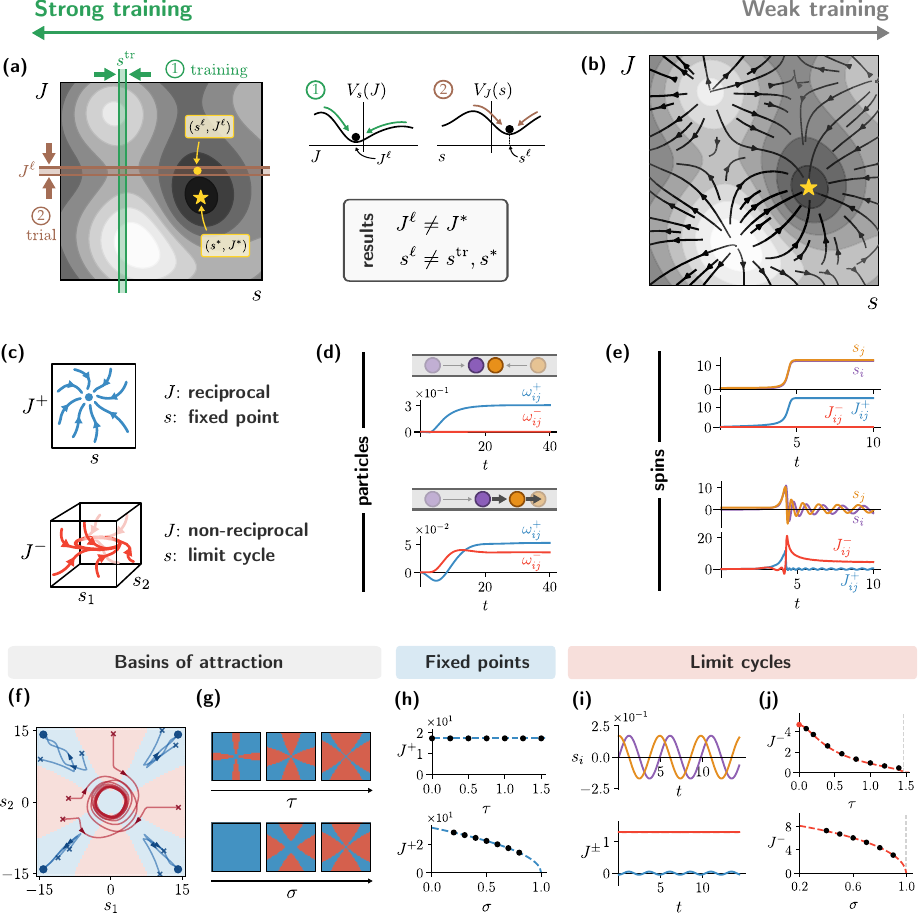}
\caption{\textbf{Learning on the fly.}
(a) Schematic understanding of learning through external training. In a (fictitious) energy landscape for the system's state $s$ and interactions $J$, training fixes the state $s^{\text{tr}}$ while the interactions find an effective minimum at $J^{\ell}$. Then, the retrieval fixes $J^{\ell}$ while the state evolves to an effective minimum at $s^{\ell}$. Generically, this procedure cannot reproduce the training behaviour exactly, i.e. $s^{\text{tr}}\neq s^{\ell}$.
(b) Schematic understanding of learning \textit{without} external training.
(c) The dynamics can lead to fixed points in reciprocal $J^+$ and $s$ (top), or fixed points in non-reciprocal $J^-$ with limit cycle steady states in $\mb{s}$ (bottom). (d) Evidence from simulation for self-learning fixed points (top) and limit cycles (bottom) in the chemotactic particle system. (e) Analogous simulation results in the spin system.
(f)-(j) Self-learning in the system of two spins. (f) Trajectories for the system's evolution in configuration space. Basins of attraction for fixed point (limit cycle) solutions are shaded blue (red). (g) Basins of attraction as a function of parameters describing STDP, $\sigma$ and $\tau$. 
(h) Analytical calculations of fixed point solutions (dashed lines) with numerical validation (solid points). (i) Simulation data of $s_i$ and $J^\pm$ at steady state in the limit cycle case. (j) Limit cycle solutions (dashed lines are analytical, solid points from simulation.) }
\label{fig4}
\end{figure*}

\medskip
\textit{Time-delayed learning rule from causal propagation.}\textemdash 
We now integrate out the concentration field $c_j(\mb{r},t)$ to obtain a closed-form dynamics for the mobilities $\omega_{i j}$
in terms of the (causal) Green function $G$ of Eq.~\eqref{dot_c}.
In the limit where $\Theta(|\mathbf{r}-\mathbf{r}_j(t)| - \ell_c) \simeq \delta(\mathbf{r}-\mathbf{r}_j(t))$, the learning rule Eq.~\eqref{dot_om} becomes
\begin{align}
    \frac{d{\omega}_{ij}}{d t} &= \alpha
    \bigg[
    \int_{-\infty}^{t} G(\mathbf{r}_i(t),t;\mathbf{r}_j(t'), t') dt' - c_0 
    \bigg] 
    - \frac{\omega_{ij}}{\tau_0}. \label{dot_om_ae}
\end{align}
Here a memory kernel appears explicitly.
This shows that the learning rule effectively takes into account the past state of the system $\mathbf{r}_j(t' < t)$.

Consider now a particle $i$ moving in a straight line at constant velocity $\mb{v} = v\hat{\mb{x}}$ (Fig.~\ref{fig2}a). 
In this case, the Green function in Eq.~\eqref{dot_om_ae} can be found analytically (Methods). 
As shown in Fig.~\ref{fig2}a, the corresponding comet-shaped concentration field exhibits a fore-aft asymmetry that originates from the motion, in a way similar to the oil droplets in Fig.~\ref{fig_intro}e.
As a consequence, a particle $j$ passing just behind $i$ (i.e., where $i$ was in the past) experiences a stronger signal from $i$ than if it passed just in front of it (i.e., where $i$ will be in the future), see Fig.~\ref{fig3}a.
The emergence of this time-asymmetric delay in the learning rule, which would occur with any causal (non-instantaneous) form of interaction, is the key to learn asymmetric interactions and dynamic states. 
Note that the fore-aft asymmetry of the comet-like diffusion cloud is only appreciable during training when the Péclet number $\text{Pe} = a v/D$ is high enough ($v$ is the velocity of the particles during training, and $a$ their separation).
In Methods, we show analytically that this asymmetry scales as $\sinh(\text{Pe})$ in the limit of fast decay $\lambda a^2 \gg D$. 

\medskip
\textit{Learning to self-assemble and move.}\textemdash Figure \ref{fig2}b shows that the system indeed learns asymmetric mobilities $\omega_{ij} \neq \omega_{ji}$ during training as long as $\text{Pe}$ is large enough. 
These lead to the dynamic behavior during retrieval, where particles evolve according to
Eq.~\eqref{dot_r} with the value of the interactions $\omega_{ij}$ obtained at the end of the learning phase (Fig.~\ref{fig2}c and Movie 2). 
We are now in the position of exploring what
dynamic functionalities are enabled by this learning rule. We first choose the particles to self-assemble into a ring-like structure with the additional requirement of rotating on demand in either the clockwise or counter-clockwise direction (see Movie 2 and Fig.~\ref{fig2}c). 
Note that despite the simplicity of the structure, we observe in Extended Data Fig.~\ref{fig:interaction} that the learning rule leads to a complex graph of interactions that contribute to the stability of the ring.

Next, we train the particles (violet) to learn to self-assemble into a triangular formation and collectively chase a different target particle (orange) by imposing on them examples of the motion shown in Fig.~\ref{fig2}d and Movie 2. During the retrieval period, we indeed see the particles self-assemble into a flying-wing shaped object (blue box) that cohesively maneuvers in space to chase (green box) and finally capture the moving target (red box).


As we illustrated with the LEGO toy in Movie 1, the examples in the training protocol (i.e. the motions imposed to the droplets or particles) should be time irreversible in order for the emergent time-delayed learning rule to produce asymmetric couplings.
Here, it means that, for instance, particle 1 should systematically follow particle 2, as we demonstrate in Extended Data Fig.~\ref{fig3} and Movie 3.

\medskip
\textit{Modified Hopfield spin model.}\textemdash 
In order to 
to determine the basins of attraction and domain of existence of the fixed point and limit cycle solutions encoding the learned functionalities, we introduce a minimal model in which the time-delayed learning rule is put by hand rather than emergent.
In this model, inspired by the Hopfield spin network~\cite{Hopfield1982,Ramsauer2020,Hopfield1984,Koiran1994,Parisi1986,Crisanti1987,Sompolinsky1986,Kanter1987,Kleinfeld1986,Hertz1987,Hertz1986,Clark2024}, the physical degrees of freedom are continuous spins $s_i(t) \in \mathbb{R}$ following the equation
\begin{align}
    \frac{d s_i}{d t} = \sum_{j \ne i} J_{ij} s_j - k s_i^3
    \label{dot_s}
\end{align}
in which $J_{i j}$ represent the effect of spin $j$ on spin $i$. 
We consider the Hebbian-like learning rule with a time delay~$\tau$,
\begin{align}
    \frac{d{J}_{i j}}{d t} = \alpha \,\big[
    s_i(t)s_j(t-\tau) - \sigma\,s_j(t)s_i(t-\tau)
    \big] -\kappa J_{ij}^3
    \label{dot_J}
\end{align}
in which $\sigma$ controls the degree of asymmetry in time, while $\alpha$ is a learning rate.
Insights on the robustness of the learning phenomenon can already be obtained from considering the simplest non-trivial case of two spins.

In the training phase, a sinusoidal oscillation of the two spins $s_i(t)$ is imposed, with $0$ or $\pi/2$ phase difference, corresponding to time-reversal symmetric or asymmetric training protocols respectively.
We first confirm that the asymmetric coefficient $J_{12} \neq J_{21}$ are obtained only when the training examples are time-irreversible (Extended Data Fig.~\ref{fig3}), very much like in the LEGO toy and particle based models. 
Furthermore, we can suppress $J^- = J_{12} - J_{21}$ completely by removing the delay $\tau = 0$ (Methods). 
As a consequence, both fixed points and limit cycles can be obtained in the retrieval phase (Extended Data Fig.~\ref{fig3}).

\textit{Learning on the fly.}\textemdash 
So far, we have considered physical learning as a two-step process with separate training and retrieval phases.
However, this strict distinction may be impractical in experiments.
For instance, with the oil droplets of Fig.~\ref{fig_intro}, the same physico-chemical processes are at play during training and retrieval, only with different Péclet numbers. 
Hence, the system may continue learning \enquote{on the fly} during retrieval, with potentially detrimental consequences if it ends up relaxing to an untrained state.

To assess the consequences of this intertwining, we formalize the training and retrieval process using a dynamical system perspective. The degrees of freedom in the system are split into degrees of freedom of interest $s_i(t)$ (e.g. spins) and learning degrees of freedom $J_{a}(t)$ (e.g. interactions between spins). 
The learning dynamics $\mb{\dot J} = \mb{f}_J(\mb{J},\mb{s})$ and the retrieval dynamics $\mb{\dot s} = \mb{f}_s[\mb{J},\mb{s}] + \mb{h}_{\text{ext}}$ can be seen as two \enquote{slices} of a dynamical system handling both $\mb{J}$ and $\mb{s}$ (green and brown lines in Fig.~\ref{fig4}a).
In the limit of strong training, that we considered up to now, a trajectory $\mb{s}_{\text{tr}}(t)$ is imposed by an external operator through the external field $\mb{h}_{\text{ext}}$ during the training period, leading to a learned $\mb{J^*}$ that is frozen and used during the retrieval period.
Another extreme case, that we call weak training, consists in reducing the training to setting an initial condition for $\mb{s}$, after which $\mb{J}$ and $\mb{s}$ simultaneously evolve towards the attractors of the dynamical system with $\mb{h}_{\text{ext}}=\bm{0}$ (see Fig.~\ref{fig4}), without the need of an external operator.

In Fig.~\ref{fig4}, we have overlaid a unified \enquote{learning  landscape} $V(\bm{s}, \bm{J})$ that schematically  describes the relaxation of both training $\bm{J}$ and physical degrees of freedom $\bm{s}$ towards (dynamic or static) attractors~\footnote{Note that when $\bm{f}[\bm{s},\bm{J}]$ depends on the past trajectory of the system, then $\bm{s}$ in the figure represents the full past trajectory of the system. In practice, when $\bm{f}$ does not only depend on the values of $\bm{s}(t')$ in the recent past, it approximated by the instantaneous values of $\bm{s}$, $\bm{\dot s}$, $\bm{\ddot s}$, ... up to a certain order.}.
In the case of non-potential dynamical systems, $V$ can be constructed as a Lyapunov function that captures the relaxation towards dynamic attractors such as limit cycles, but not the motion on the attractors~\cite{Fang2019}.
As shown in Fig.~\ref{fig4}b, we expect that the strong learning that we used up to now does not necessarily lead to minima of the learning landscape: the state obtained in the free evolution after learning $\bm{s}^\ell$ may not exactly reproduce the training state $\bm{s}^{\text{tr}}$ (Fig. \ref{fig4}b). 

This is indeed what we observe, in the particle system (see Methods, in which we also describe the effect of a finite training time) as well as in the minimal spin model with $N=2$ where exact solutions can be obtained (see Methods and Fig. \ref{fig4}h-j).
Here, the training is reduced to the bare minimum of an initial condition for $\mb{s}$, while we set $\mb{J}=\mb{0}$ at the initial time $t=0$.
Figure ~\ref{fig4}f shows the results of our calculations of the basins of attraction of the fixed point (in blue) and limit cycles (in red): sufficiently asymmetric initial conditions spiral towards limit cycles, while more symmetric ones converge to fixed points. 
In the Methods, we show that an exact solution of the minimal model in Eq. 6 allows us to prove that the size of the basins of attraction is controlled by the degree of asymmetry $\sigma$ and the time delay $\tau$ of the time-delayed learning rule (Fig. \ref{fig4}g). 
In addition, limit cycles disappear at critical values $\sigma_c < 1$ and $\tau_c > 0$ (Fig. \ref{fig4}h-j).
We find that, when the time delay $\tau$ is larger than the critical value $\tau_c$, asymmetric interactions are suppressed before they can be learned.
This suggests that there is a trade-off between large basins of attraction (at large $\tau$) and large non-reciprocity (at small but finite $\tau$).

To sum up, when the interactions are not frozen at the end of training, the system evolves towards its natural attractors, potentially leading to undesired configurations. However, the analysis behind Fig.\ref{fig4} reveals that one can also harness this feature as a shortcut to reduce the training time so that the evolution of the interactions $\mb{J}$ and of the physical degrees of freedom $\mb{s}$ reinforce each other.

\textit{Conclusions.}\textemdash 
We have shown how to train and retrieve dynamic states through physical learning by combining time-delayed interactions, that emerge naturally from causal propagation, with a time-reversal asymmetric training protocol.
This strategy can not only be applied in robotic matter by directly programming a computer, but also in physico-chemical active matter systems where it naturally emerges. 
Looking forward, programmable colloids based on DNA and other biomolecules~\cite{stern_learning_2023,Fredriksson2002,Schaus2017,Moerman2023,Weinstein2019,Owen2023} would greatly enhance the potential complexity of learned behaviors by providing multiple interacting species and chemical feedback loops through enzymatic reactions. 
The principles illustrated here provide a step towards von Neumann's dream of engineering synthetic living systems that adapt to the environment.


\clearpage

\section*{Materials and Methods}

\renewcommand{\figurename}{Extended Data Figure}
\renewcommand{\tablename}{Extended Data Table}
\setcounter{figure}{0}

\subsection*{Proposal of experimental realization using chemotactic oil droplets}

\begin{figure*}[h!]
{\centering
\includegraphics[width=0.9\textwidth]{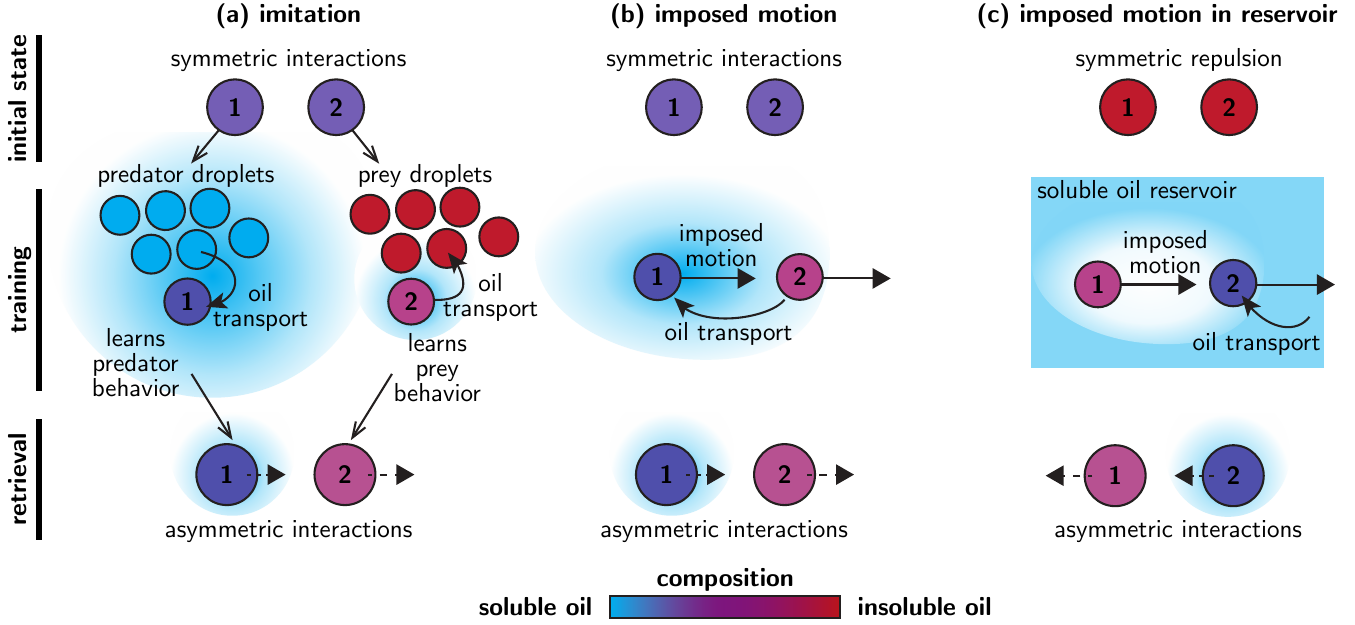}}
\caption{\label{EDF_oil}
\textbf{Proposed training mechanisms for oil droplets to learn asymmetric interactions.} In panel (a) the training consists of spending time near either a predator or a prey droplets. In panel (b), the training consists of moving the droplets in a pair with some external force (optical tweezers, electric field). In panel (c), the droplets are also moved but in a reservoir of soluble oil, causing the droplets to move during the retrieval phase in the opposite direction from which they moved in the training phase.
}
\end{figure*}

In this section, we discuss how a system of oil droplets that learn asymmetric interactions (such as that in Fig.~\ref{fig_intro}e-f of the main text) can be realized experimentally. 
We suggest a recipe based on the chemotactic chasers from Ref.~\cite{Meredith2020} but expect that learning rules with time delays could be implemented in other situations as long as there is an exergonic process coupled to the objects’ motion that can drive their dynamics. 

The chasing droplets in Ref.~\cite{Meredith2020} have asymmetric interactions that stem from asymmetric oil exchange. Two types of droplets, each consisting of an oil that is fully miscible with the oil in the other droplet, exchange oils through the aqueous continuous phase via micellar solubilization. One of the two oils is more soluble in the continuous phase than the other, so that the oil transport of droplet 1 to droplet 2 is faster than the transport of droplet 2 to droplet 1. The consequence is a net loss of soluble oil by droplet 1 and a net gain of soluble oil by droplet 2. This directional oil exchange leads to motion because the droplets tend to move away from areas with high solubilized oil concentrations; the solubilized oil increase the interfacial tension of the oil droplets driving a Marangoni flow. The droplets made of soluble oil thus moves toward the droplets made of the insoluble oil and, conversely, the droplet made of insoluble oil moves away from the droplet made of soluble oil. The net result is an asymmetric interaction where the soluble oil droplets act as predators and chase the insoluble oil droplets that act as prey.

The predator and prey droplets have an inherent memory effect through their composition; their oil exchange rate depends on their local chemical environment: droplets near predators lose less soluble oil, and droplets near prey lose more soluble oil. The droplets `remember', through their composition, how many predator or prey droplets they have been near on average. 

This memory effect could be used to have droplets of initially identical composition---and thus symmetric interactions---develop asymmetric interactions over time by evolving their oil exchange process in differing environments (Extended Data Fig.~\ref{EDF_oil}a). A mixed-composition droplet that spends time near mostly predator droplets loses less soluble oil (or may even gain soluble oil depending on its initial composition) so that its composition becomes more predator-like. Conversely, a mixed-composition droplet that spends time near mostly prey droplets loses more soluble oil and becomes more prey-like. When removed from their training chemical environment and placed together, the droplet that has evolved to a predator-like composition would likely chase the droplet that has evolved to a prey-like composition. 

The same physical process of chemical-environment-dependent oil exchange could lead to droplets learning asymmetric interactions by the act of imposing the droplet motion in which the asymmetric interactions will result (Extended Data Fig.~\ref{EDF_oil}b). Imagine two droplets of equal mixed oil composition are forced to move in a pair with one droplet in front and the other at the back. When this motion if fast compared to diffusive transport (i.e. the Péclet number is much larger than one), the droplet in the back experiences a chemical environment with more soluble oil than the droplet in front. The back droplet can thus uptake oil from the front droplet but not the other way around, which causes the front droplet to have less soluble oil and become more prey-like, and the back droplet to have more soluble oil and become more predator-like. 

Note that in the two cases described above, the training would be mechanistically similar to the retrieval. However, this is not necessarily the case: for instance, we expect that the same training protocol performed in a reservoir of soluble oil would cause the droplets to move in opposite directions during training and retrieval (Extended Data Fig.~\ref{EDF_oil}c).

\begin{figure*}[]
\centering
\hspace{-1em}\includegraphics[width = 1.2\columnwidth]{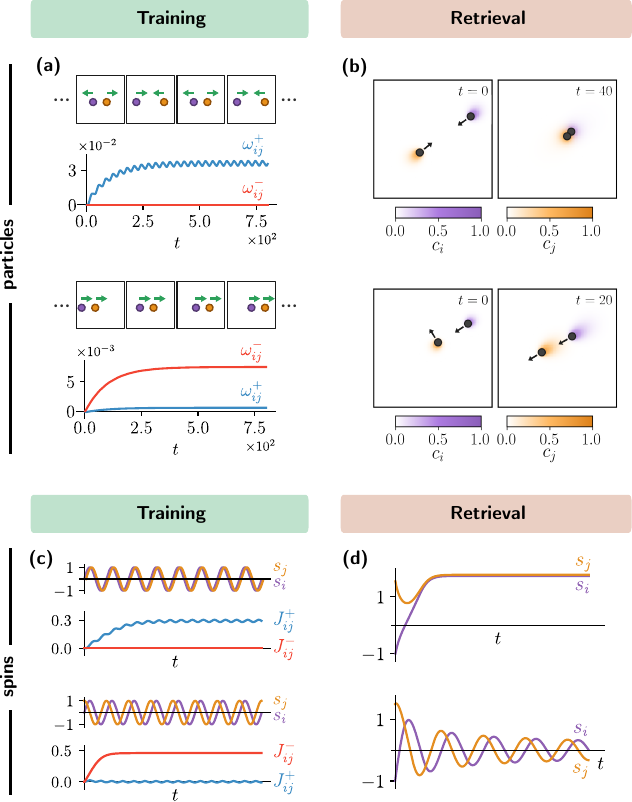}
\caption{\textbf{Dynamic steady states require both time delay in the learning rule and broken time-reversal invariance (TRI).} 
(a) Simulation data for the evolution of $\omega_{ij}^\pm$ during training show symmetric interactions being learned (top) when the training protocol is time reversal invariant and asymmetric interactions emerging (bottom) where $i$ consistently follows $j$ (TRI broken). 
(b) Snapshots showing retrieval periods for the TRI (top) and TRI broken (bottom) training protocols.
(c) The spins $s_i$ learn symmetric interactions during the training protocol with sinusoidal oscillation and no phase difference, and they learn asymmetric coefficients when the phase difference is $\pi/2$. (d) During the retrieval period spins with symmetric interactions reach an aligned, stationary steady state whereas the ones with asymmetric interactions show oscillatory steady state. Colors match the labels in 
(c).
}
\label{fig3}
\end{figure*}

\subsection*{Demonstration with LEGO toys}

In Movie 1, we illustrate the ingredients required to learn dynamical states using a programmable toy.
Two \enquote{Technic Medium Angular Motor} from a \enquote{LEGO Education SPIKE Prime} set are used, both as motors and angle sensors, to realize active XY-like spins.
The LEGO toy is programmed to approximate the dynamics
\begin{equation}
\dot{\theta}_i = f_i(\vec{\theta}) \equiv \sum_{j} J_{i j} \sin(\theta_j - \theta_i)
\end{equation}
in which $i,j=1,2$ label the XY spins with angles $\theta_i(t)$ (see Ref.~\cite{Fruchart2021} for an exact solution of this dynamical system).
This is realized by moving the positions of the robots by an increment $f_i(\vec{\theta}) \delta t$ every time step.
In order to allow the user to move the spins by hand during the operation, the motors are also paused for a fixed time $\tau_{\text{pause}}$ every time step.

In the training phase, the motors are disabled and time series of the angular positions $\theta_i$ imposed by the operator are recorded.
The couplings are learned from these time series as
\begin{equation}
\dot{J}_{i j} = J_{i j} - \kappa J_{i j}^3 + \alpha \int_{-\infty}^{t} g(t-t') \cos(\theta_1(t) - \theta_2(t')) \text{d}t.
\end{equation}
in which the memory kernel is taken as $g(\tau) = \text{e}^{-\tau/\tau_0}$ (and is set to zero for times going before the beginning of the recording).
The parameters $\kappa$, $\alpha$, and $\tau_0$ are adjusted to be in a regime where learning is successful (see the main text for an analysis of the influence of training parameters).

\subsection*{Agent-based particle model}

\subsubsection{Simulation details}

As mentioned in the main text, for the particle based simulation we use overdamped dynamics of self-propelled particles following,
\begin{equation}
\frac{d{\mathbf{r}}_i}{dt}=v_0 \mathbf{n}_i - \sum_{j\ne i}\nabla_i V_{ij} +{\boldsymbol{\xi}}(t)
\end{equation}
where $\mathbf{r}_i$ is the position of the $i$ the particle, $v_0$ is the propulsion speed which acts along the direction $\mathbf{n}_i=(\cos{\theta_i},\sin{\theta_i})$. Here $V_{ij}$ is a short range repulsive power law potential of the form
\begin{equation}
V_{ij}(r)=
\begin{cases}
    \epsilon \(\frac{\sigma}{r}\)^{10} + V_0 + \frac{1}{2} V_2 r^2 & \text{if } r \leq r_c\\
    0              & \text{otherwise}
\end{cases}
\end{equation}
where $\epsilon$ and $\sigma$ determines the energy scale and length scale of this steric interaction and $r_c$ is the cutoff distance. We have chosen $\epsilon=1.0$, $\sigma=1.0$, and $r_c= 1.05 \sigma$ throughout our simulation. Also we have added the constant and the quadratic term in the potential to make both the potential and the corresponding force continuous at $r=r_c$ by choosing $V_0=-6 \epsilon \left(\frac{\sigma}{r_c}\right)^{10}$ and $V_2=10 \epsilon \left(\frac{\sigma^{10}}{{r_c}^{12}}\right)$. In the equation of motion mentioned above we used a small translational noise ${\boldsymbol{\xi}}$ whose components are uniformly distributed between $[-\eta_t, \eta_t]$. The trained coefficients $\{\omega_{ij}\}$ are used to determine the direction of self-propulsion $\mathbf{n}_i \equiv (\cos{\theta_i},\sin{\theta_i})$ mentioned before. The equation of motion for $\theta_i$ is
\begin{equation}
\frac{d \theta_i(t)}{dt}=\frac{\sin{\(\theta^T_i(t)- \theta_i(t)\)}}{\tau_{\theta}} + 2 \pi \xi_r
\label{tr2}
\end{equation}
where $\theta_i^T(t)$ represents the instantaneous target angle for the $i$-th particle and it reflects the orientation of the vector $\sum_{j\ne i} \omega_{ij} \nabla c_j(\mathbf{r}_i,t)$. Note that this introduces a finite time relaxation mechanism in the orientational dynamics and in the limit $\tau_{\theta} \to 0$, we get $\mathbf{n}_i \to \tilde{\mathbf{n}}_i$ as mentioned in the main text. The timescale for the relaxation dynamics is controlled by $\tau_{\theta}$ and $\xi_r$ is a rotational noise which is uniformly distributed between $[-\eta_r, \eta_r]$. In the trial phase the system is initialised with randomised configuration of particles and zero pheromone concentration in the box. Particles then start releasing the pheromones (following Eq.~\ref{dot_c} of the main text) and have constant self propulsive overdamped dynamics with speed $v_0$ which try to align towards a target angle $\theta_i^T(t)$, determined by the trained affinities $\omega_{ij}$.

\subsubsection{Order parameters}

To identify the nature of the dynamic steady state observed during the trial in the particle based simulation, we construct the following order parameter for a pair ($i,j$)
\begin{equation}
\eta_{ij} = \tfrac{1}{2}(\hat{\mathbf{n}}_i+\hat{\mathbf{n}}_j)\cdot \hat{\mathbf{r}}_{ij}
\end{equation}
where $\hat{\mathbf{n}}_i$ is the unit vector associated with the propulsion vector of the $i$ th particle and $\hat{\mathbf{r}}_{ij}$ is the unit vector along the line joining the pair ($i,j$). This order parameter by construction will be $\eta_{ij} \sim$ zero for reciprocal attraction or repulsion as the center of mass velocity is zero when particle $i$ and $j$ either approach or run away from each other. However for a chase and run scenario, $ \tfrac{1}{2}(\hat{\mathbf{n}}_i+\hat{\mathbf{n}}_j) \sim \hat{\mathbf{r}}_{ij}$ and the order parameter will be $\sim 1$.

\begin{figure*}
\centering
\includegraphics[height =.4\linewidth]{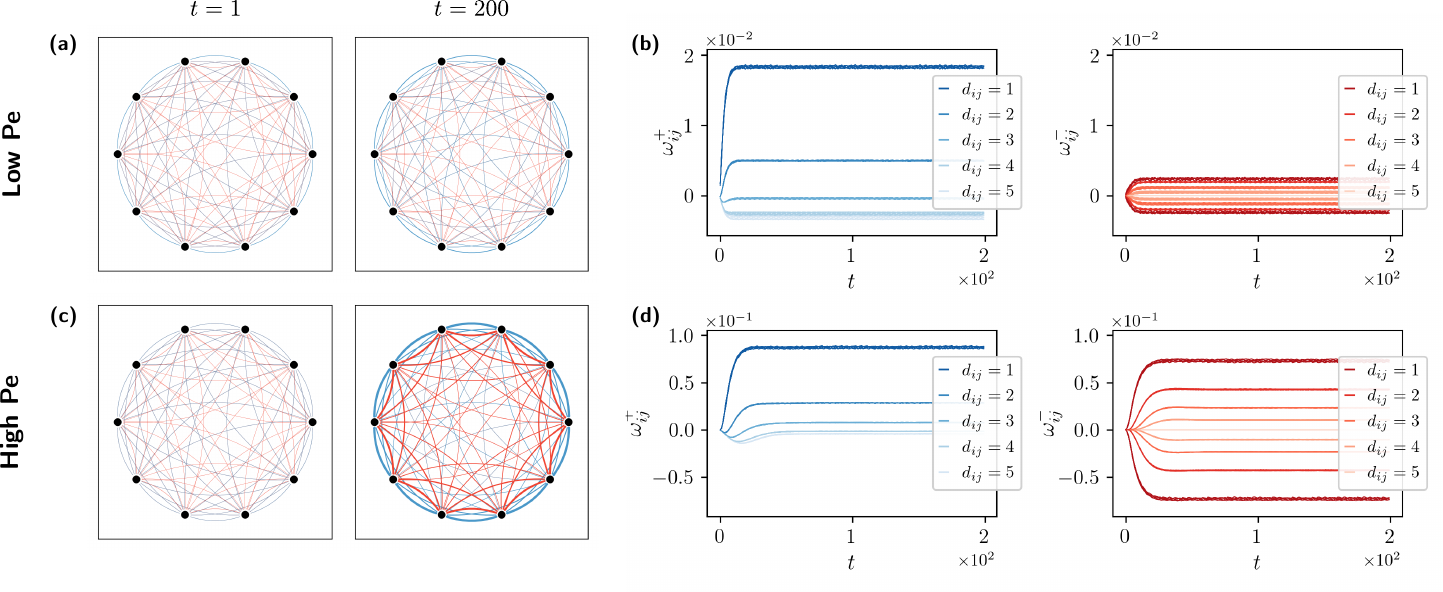}
\caption{{\bf{Evolution of interaction during training.}} (a) Pair-wise interaction map in the beginning (left) and at the end (right) of the training phase for low P\'eclet number. The symmetric part of the interaction $\omega_{ij}^+$ is plotted in blue and the anti-symmetric part of the interaction $\omega_{ij}^-$ is plotted in red. (b) Both the symmetric and the anti-symmetric part of the interactions ($\omega^{+}_{ij}$) and ($\omega^{-}_{ij}$) are plotted as a function of time $t$ during the training for low P\'eclet number. The interactions are colored according to the distance between the pairs ($i$ and $j$) where $d_{ij}=|i-j|$. (c) and (d) are analogous plots to (a) and (b) for high P\'eclet number. The emergent interaction is more asymmetric (the network has more red connections) when the P\'eclet number is high (bottom row) than when the P\'eclet number is low (upper row). Note that the symmetry in (d) arises because $\omega_{ij}^- = -\omega_{ji}^-$.}
\label{fig:interaction}
\end{figure*}

\subsection*{Field created by a particle moving at constant speed}

In this section, we compute the steady-state field profile in the co-moving frame of reference of a particle moving at constant velocity $v_0$. 
In this case, $c(\mathbf{r},t)$ follows the equation
\begin{equation}
\frac{\partial c(\mathbf{r},t)}{\partial t}={\tau_c}^{-1} \delta(\mathbf{r}-\mathbf{r}_0(t))-\lambda c(\mathbf{r},t)+ D \nabla^2 c(\mathbf{r},t)
\label{phevol}
\end{equation}
in which $\mathbf{r}_0(t) = v_0 t$ is the position of the particle (we set $\mathbf{r}_0(0) = \bm{0}$). In the co-moving frame of reference, we define two coordinates $u=x-v_0 t$ and $w=y$.

A Fourier transform of Eq.~\eqref{phevol} yields 
\begin{equation}
\tilde{c}(q_u,q_w)=\frac{1}{2 \pi \tau_c \(\lambda -i v_0 q_u+ D({q_u}^2+{q_w}^2) \)}
\end{equation}
where $(q_u,q_w)$ is the spatial frequency.
The inverse Fourier transform is evaluated using the residue theorem 
and we end up with
\begin{equation}
c(u,w)=\frac{1}{4 \pi^2 D \tau_c} \, I(u,w)
\end{equation}
in which 
\begin{equation}
I(u,w)=2 \pi \, e^{-u/\xi} \, K_0[\zeta^{-1}(u^2+w^2)^{1/2}]
\end{equation}
where we have defined $\xi^{-1}={v_0}/{2 D}$ and $\zeta^{-2}=\xi^{-2}+{\lambda}/{D}$ and where $K_\nu$ is the modified Bessel function of the second kind of order $\nu$ \cite[\href{https://dlmf.nist.gov/10.25}{\S 10.25}]{NIST_DLMF}.

In the limit $w \to 0$, an asymptotic expansion of the Bessel function~\cite[\href{https://dlmf.nist.gov/10.25\#E3}{\S 10.25.3}]{NIST_DLMF} yields
\begin{equation}
c(u,w \to 0)=\frac{1}{2 D \tau_c} \sqrt{\frac{\zeta}{2 \pi |u|}} e^{-\left(\frac{1}{\zeta} + \text{sign}(u) \frac{1}{\xi}\right)|u|}.
\label{prof}
\end{equation}
We can then quantify the fore-aft asymmetry by computing
\begin{equation}
    c(-a) - c(a) = \frac{1}{D \tau_c} \sqrt{\frac{\zeta}{2 \pi a}} 
    e^{-a/\zeta}
    \sinh(a/\xi)
\end{equation}
for $a > 0$, in which we identify $a/\xi = v_0 a/[2D] = \text{Pe}_{a/2}$ as the Péclet number associated with the length $a/2$.

\subsection*{Modified Hopfield spin model}

The model is defined by Eqs.~\eqref{dot_s} and \eqref{dot_J} of the main text, for $i,j=1,\dots,N$. It is convenient to introduce the symmetric and antisymmetric parts of the coupling $J_{ij}^\pm = ({J}_{ij} \pm {J}_{ji})/2$ (for $i \geq j$, as the others $J_{ij}^\pm$ are redundant).

\subsubsection{Role of delay time $\tau$}
In this section, we elucidate how the time delay in the learning rule enables the emergence of asymmetric interactions during training. Fig. \ref{fig3} compares the outcome when the oscillatory training protocol of Fig. \ref{fig3}q-x is performed on the spin system with delay times $\tau = 0$ (non-STDP) and $\tau \neq 0$ (STDP). When the training protocol is time-reversal invariant (Fig. \ref{fig3}a), neither case results in non-reciprocity. However, when the training protocol breaks time-reversal invariance (Fig. \ref{fig3}b), it is only in the STDP case $\tau \neq 0$ that results in $J_{ij}^- \neq 0$ after training. The non-STDP case has $J_{ij}^- = 0$ throughout the entire training period, even as $J_{ij}^+$ oscillates around zero.

Indeed,  when $\tau = 0$, the equation of motion for $J_{ij}^- = ({J}_{ij} - {J}_{ji})/2$ obtained from Eq.~\eqref{dot_J} becomes
\begin{equation}
\begin{split}
    \label{s2-dot_J-}
    \dot{J}_{ij}^- = -\kappa J_{ij}^- \big[ 3(J_{ij}^+)^2 + (J_{ij}^-)^2 \big] 
\end{split}
\end{equation}
so $J_{ij}^-$ can only monotonically decay to zero.
In contrast, $J_{ij}^+$ may remain finite.


\subsubsection{Weak learning: exact solution for $N=2$.}
The dynamical system composed of spins $s_1$, $s_2$ and their interactions $J_{12}$ and $J_{21}$ (or equivalently $J^\pm = (J_{12}\pm J_{21})/2 $) has fixed point and limit cycle steady states, as shown in Fig. \ref{fig4}b-d. Here, we exactly solve for these steady states, in particular to understand how they depend on the STDP parameters $\tau$ and $\sigma$. 
We have
\begin{subequations}
\begin{align}
   \dot{s}_1 &= (J^+ + J^-) s_2 - k s_1^3 \\
   \dot{s}_2 &= (J^+ - J^-) s_1 - k s_2^3 
\label{s2-dot_s}
\end{align}
and 
\begin{equation}
\label{s2-dot_J+_2}
\begin{split}
    \dot{J}^+ =
    \frac{\alpha}{2} \,(1-\sigma)\,\big[
    s_1(t)s_2(t-\tau) + \,s_2(t)s_1(t-\tau)
    \big]  \\
    -\kappa J^+ \big[ 3(J^-)^2 + (J^+)^2 \big] 
\end{split}
\end{equation}
and
\begin{equation}
\begin{split}
    \label{s2-dot_J-_2}
    \dot{J}^- =
    \frac{\alpha}{2} \,(1+\sigma)\,\big[
    s_1(t)s_2(t-\tau) - \,s_2(t)s_1(t-\tau)
    \big] \\
    -\kappa J^- \big[ 3(J^+)^2 + (J^-)^2 \big] 
\end{split}
\end{equation}
\end{subequations}

\paragraph{Fixed points.}
Fixed points are obtained by setting $\dot{s}_1 = \dot{s}_2 = \dot{J}^+ = \dot{J}^- = 0$ (note that this implies $s_i(t) = s_i(t-\tau)$).
We obtain
\begin{subequations}
\begin{align}
    s_i &= \pm \sqrt{\tfrac{\abs{J^+}}{k}}\\
    J^+ &= \text{sign}(s_1 s_2) \sqrt{\tfrac{\alpha}{k\kappa}(1-\sigma)}
    \\
    J^{-} &= 0
\end{align}
\end{subequations}
where the signs of $s_1$ and $s_2$ are independent.

\paragraph{Limit cycles.}
To find limit cycle solutions, we set $\dot{J}^- = 0$ (to match numerical observations) and assume that $\abs{J^+} \ll \abs{J^-}$ (we will later derive conditions for this to be true). In this case, we neglect $J^+$ in Eq.~(\ref{s2-dot_s}) and obtain
\begin{subequations}
\label{s2-sol_s_i}
\begin{align}
   s_1(t) = s \, \cos(J^- t) \\
   s_2(t) = s \, \sin(J^- t) 
\end{align}
\end{subequations}
where we assumed $s^2 \ll J^-$. We will solve for $s$ and derive conditions for this as as well. \\

Next, the solutions for $s_i$ can be placed into Eq. (\ref{s2-dot_J+_2}) and (\ref{s2-dot_J-_2}) to find $J^\pm$. With $\dot{J}^- = 0$, Eq. (\ref{s2-dot_J-_2}) the relation between $J^-$ and s
\begin{align}
    \frac{2\kappa}{\alpha(1+\sigma)} = \frac{s^2 \sin(J^- \tau)}{(J^-)^2} \label{s2-sol_J-_1}.
\end{align}
while Eq. (\ref{s2-dot_J+_2}) can be integrated to get
\begin{align}
     J^+(t) = -J^+\cos{\left( 2J^-t - J^-\tau + \phi \right)}, \quad\quad 
.\label{s2-sol_J+}
\end{align}
with 
\begin{align}
     J^+ &= \kappa \left(\frac{1-\sigma}{1+\sigma}\right) \left(\frac{(J^-)^2}{\sin(J^- t)}\right) \left(4+ 9\kappa^2(J^-)^2\right)^{-1/2}\\
     \phi &= \tan^{-1}\left(\frac{3}{2} \kappa J^- \right)
\end{align}
We see that the initial assumption of $\abs{J^+(t)} \ll \abs{J^-}$ is valid when $\sigma$ is sufficiently close to 1, corresponding to the antisymmetric limit of STDP.\\

The steady state value of the amplitude $s$ is obtain using the above solutions for $s_i$ and $J^\pm$ as follows:
\begin{align}
    s\frac{ds}{dt} &=  s_1\frac{ds_1}{dt} + s_2\frac{ds_2}{dt} \nonumber\\
    &= -s^2 J^+\, \sin(2J^- t)\cos{\left( 2J^-t - J^-\tau + \phi \right)} \nonumber
    \\
    & \qquad - k s^4 \left(\sin^4(J^- t) + \cos^4(J^- t)\right) \nonumber.
\end{align}
Averaging over a cycle of time $2\pi/J^-$ (with $\langle\rangle$ denoting cycle average),
\begin{align}
    \left\langle\frac{ds}{dt}\right\rangle = \frac{1}{2} s J^+ \sin{\left(- J^-\tau + \phi \right)} - \frac{3}{4}k s^3 \nonumber.
\end{align}
Note that $J^+$ can no longer be ignored since the contribution of $J^-$ vanishes. Then, the steady state solution $\left\langle ds/dt\right\rangle = 0$ yields
\begin{align}
    s^2 = \frac{2 J^+\sin{\left(- J^-\tau + \phi \right)}}{3k} . \label{s2-sol_s}
\end{align}
Similar to the condition on $J^+$, the assumption $s^2 \ll J^-$ is valid for $\sigma$ close to 1.\\

Finally, we can use the solutions for $s$ (\ref{s2-sol_s}) and $J^+$ (\ref{s2-sol_J+}) to eliminate these variables in Eq. (\ref{s2-sol_J-_1}). The resulting equation reads
\begin{align}
    \frac{3k}{\alpha(1-\sigma)} = \frac{\sin{\left( - J^-\tau + \phi \right)}}{J^- \left(4+ 9\kappa^2(J^-)^2\right)^{-1/2}} \label{s2-sol_J-}
\end{align}
and can be solved numerically for $J^-$. Note that the solutions $\pm J^-$ are identical upon interchanging spin indices 1 and 2. \\

Together, Eqs. (\ref{s2-sol_s_i}), (\ref{s2-sol_J+}), (\ref{s2-sol_s}), and (\ref{s2-sol_J-}) constitute the limit cycle solutions of the dynamical system as discussed in the text and shown in Fig. \ref{fig4}(d)(ii). We now derive the condition for existence of solutions that establishes the critical values $\sigma = \sigma_c$ and $\tau = \tau_c$. The left side of Eq. (\ref{s2-sol_J-}) is maximized at $J^- \rightarrow 0$, implying the condition
\begin{align}
    \alpha \left( \frac{3}{2}\kappa - \tau_c \right) > \frac{6k}{(1-\sigma_c)}
\end{align}
for limit cycle solutions to exist. Solutions vanish as $\sigma$ or $\tau$ increase past the critical values.

\subsection{Sloppy trainer}

In the main text we formulated the problem in such a way that during the training the trajectory of the learning particles were either perfectly micro-managed by the trainer or left completely free during the self learning. We now ask, what happens if the trainer is sloppy {\it{i.e.}} the scenario is between these two extremes. 

\begin{figure*}
\centering
\includegraphics[width = 1.1 \columnwidth]{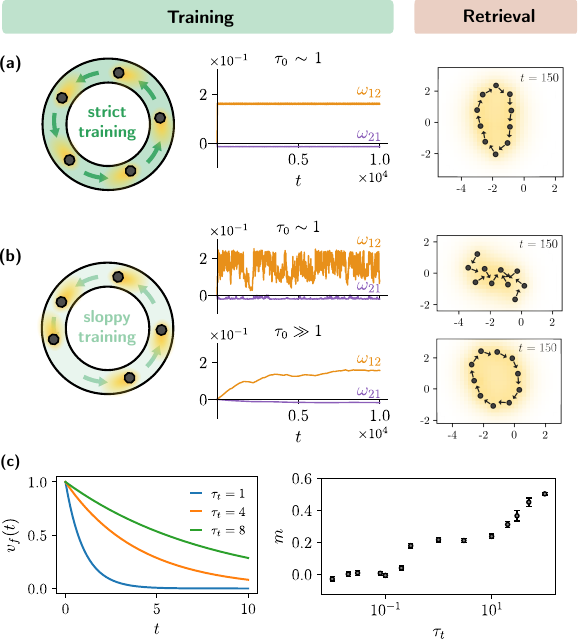}
\caption{{\bf{Sloppy trainer.}} (a) Schematic of the set up for strict training where the inter-particle distances are maintained precisely at a fixed value during the training. The evolution of the inter-particle interaction for the pair 1-2, ($\omega_{12}$ and $\omega_{21}$) shows a steady value and using these learned interactions the system successfully retrieves the circular motion. (b) During ``sloppy training", the role of trainer is implemented via a confining harmonic potential and a constant tangential flow field ($\mathbf{v_f}(t)=v_f(t) \hat{\phi}$ and $\hat{\phi}$ is the unit vector along the azimuthal direction). In the case of the sloppy trainer, the interparticle distances can fluctuate and this leads to strong fluctuations in the learned interactions, which prevents a successful retrieval of the behavior. However, if the learning is slow ($\tau_0>>1$), the learning process averages out the fluctuation and a successful retrieval is possible. (c) (Left) Temporal sloppiness in the training procedure is implemented by a exponentially declining velocity field $v_f(t)=\exp(-t/\tau_t)$. (Right) The performance of the particles improves with increasing $\tau_t$, especially after crossing a threshold at $\tau_t \sim 1$ which is depicted by a dynamic order parameter $m$ (see Materials and Methods: ``Sloppy trainer" for details).}
\label{fig:sloppy}
\end{figure*}

We address this in two different ways: (a) by reducing the spatial control and (b) by decreasing temporal one. To understand the role of sloppiness of the trainer, we consider that the particles are confined in an annular domain (which acts as a quasi-linear track with harmonic confinement with stiffness $\kappa=100$) during the training. In this set up, the particles are supposed to perform a periodic circular motion during the training. The presence of the trainer is implemented in the equation of motion of the learning particles via a tangential flow field $\mathbf{v}_f(t)=v_f(t) \hat{\phi}$ (here $\hat{\phi}$ is the unit vector along the azimuthal direction) along with a uniformly distributed translational noise, 
\begin{equation}
\dot{\mathbf{r}}_i=v_f(t) \hat{\boldsymbol{\phi}} - \sum_{j\ne i}\nabla_i V_{ij} +\boldsymbol{\xi}_t
\label{sloppy}
\end{equation}
For the first part, {\it{i.e.}} for lack of spatial control, we use a time independent flow where $v_f(t)=v_f$. One can see that this training protocol does not preserve the relative distance between the particles and therefore due to the fluctuations in the inter-particle distance the learned coefficients shows giant fluctuations compared to the case where the training is carried out by a strict trainer (see Fig.~\ref{fig:sloppy}(a)). This fluctuations lead to set of learned ${\omega_{ij}}$ that is insufficient to retrieve the assembly where the system was trained. The particles can still learn meaningful interactions during the learning process by ramping up the time scale associated with learning or $\tau_0$ (see Eq.3 of the main text). In the last panel of Fig.~\ref{fig:sloppy}(a) one can see the coefficients ${\omega_{ij}}$ reach similar values as compared to the case which is carried out by a strict trainer. This suggests when the trainer is sloppy a slow learner (with larger $\tau_0$) is a better learner.

Next, to understand the role of temporal control, the flow field is slowly withdrawn (in an exponential manner) with a timescale $\tau$ (see  Fig.~\ref{fig:sloppy}(b)). In this case we combine both training and retrieval into a single process where the particles keeps on learning on the fly and interact via the instantaneous values learned interaction. Now the question we ask is, how large $\tau$ has to be so that the particles learn to systematically circle around in the annular chamber. Fig.~\ref{fig:sloppy}(c) shows if $\tau<1$  particles don't learn anything meaningful or $m \sim 0$ (measured by a dynamic order parameter $m=\sum_i \mathbf{v}_i(t) \cdot \hat{\phi}(\mathbf{r}_i(t))$ where $\hat{\phi}(\mathbf{r})$ is the unit vector along the azimuthal direction measured at position $\mathbf{r}$ at the steady state) whereas with $\tau>1$ the performance during the trial  is significant ($m \sim 1$).

\subsection{Movies}

Supplementary movies are available in the following web link: {\url{https://home.uchicago.edu/~vitelli/videos.html}}.

\bibliography{references}

\begin{thebibliography}{71}%
\makeatletter
\providecommand \@ifxundefined [1]{%
 \@ifx{#1\undefined}
}%
\providecommand \@ifnum [1]{%
 \ifnum #1\expandafter \@firstoftwo
 \else \expandafter \@secondoftwo
 \fi
}%
\providecommand \@ifx [1]{%
 \ifx #1\expandafter \@firstoftwo
 \else \expandafter \@secondoftwo
 \fi
}%
\providecommand \natexlab [1]{#1}%
\providecommand \enquote  [1]{``#1''}%
\providecommand \bibnamefont  [1]{#1}%
\providecommand \bibfnamefont [1]{#1}%
\providecommand \citenamefont [1]{#1}%
\providecommand \href@noop [0]{\@secondoftwo}%
\providecommand \href [0]{\begingroup \@sanitize@url \@href}%
\providecommand \@href[1]{\@@startlink{#1}\@@href}%
\providecommand \@@href[1]{\endgroup#1\@@endlink}%
\providecommand \@sanitize@url [0]{\catcode `\\12\catcode `\$12\catcode `\&12\catcode `\#12\catcode `\^12\catcode `\_12\catcode `\%12\relax}%
\providecommand \@@startlink[1]{}%
\providecommand \@@endlink[0]{}%
\providecommand \url  [0]{\begingroup\@sanitize@url \@url }%
\providecommand \@url [1]{\endgroup\@href {#1}{\urlprefix }}%
\providecommand \urlprefix  [0]{URL }%
\providecommand \Eprint [0]{\href }%
\providecommand \doibase [0]{https://doi.org/}%
\providecommand \selectlanguage [0]{\@gobble}%
\providecommand \bibinfo  [0]{\@secondoftwo}%
\providecommand \bibfield  [0]{\@secondoftwo}%
\providecommand \translation [1]{[#1]}%
\providecommand \BibitemOpen [0]{}%
\providecommand \bibitemStop [0]{}%
\providecommand \bibitemNoStop [0]{.\EOS\space}%
\providecommand \EOS [0]{\spacefactor3000\relax}%
\providecommand \BibitemShut  [1]{\csname bibitem#1\endcsname}%
\let\auto@bib@innerbib\@empty
\bibitem [{\citenamefont {Hebb}(2002)}]{hebb_organization_2002}%
  \BibitemOpen
  \bibfield  {author} {\bibinfo {author} {\bibfnamefont {D.~O.}\ \bibnamefont {Hebb}},\ }\href@noop {} {\emph {\bibinfo {title} {The Organization of Behavior: A Neuropsychological Theory}}}\ (\bibinfo  {publisher} {{L. Erlbaum Associates}},\ \bibinfo {address} {{Mahwah, N.J}},\ \bibinfo {year} {2002})\BibitemShut {NoStop}%
\bibitem [{\citenamefont {Stern}\ and\ \citenamefont {Murugan}(2023)}]{stern_learning_2023}%
  \BibitemOpen
  \bibfield  {author} {\bibinfo {author} {\bibfnamefont {M.}~\bibnamefont {Stern}}\ and\ \bibinfo {author} {\bibfnamefont {A.}~\bibnamefont {Murugan}},\ }\bibfield  {title} {\bibinfo {title} {Learning {{Without Neurons}} in {{Physical Systems}}},\ }\href {https://doi.org/10.1146/annurev-conmatphys-040821-113439} {\bibfield  {journal} {\bibinfo  {journal} {Annual Review of Condensed Matter Physics}\ }\textbf {\bibinfo {volume} {14}},\ \bibinfo {pages} {417} (\bibinfo {year} {2023})}\BibitemShut {NoStop}%
\bibitem [{\citenamefont {Keim}\ \emph {et~al.}(2019)\citenamefont {Keim}, \citenamefont {Paulsen}, \citenamefont {Zeravcic}, \citenamefont {Sastry},\ and\ \citenamefont {Nagel}}]{Keim2019}%
  \BibitemOpen
  \bibfield  {author} {\bibinfo {author} {\bibfnamefont {N.~C.}\ \bibnamefont {Keim}}, \bibinfo {author} {\bibfnamefont {J.~D.}\ \bibnamefont {Paulsen}}, \bibinfo {author} {\bibfnamefont {Z.}~\bibnamefont {Zeravcic}}, \bibinfo {author} {\bibfnamefont {S.}~\bibnamefont {Sastry}},\ and\ \bibinfo {author} {\bibfnamefont {S.~R.}\ \bibnamefont {Nagel}},\ }\bibfield  {title} {\bibinfo {title} {Memory formation in matter},\ }\href {https://doi.org/10.1103/revmodphys.91.035002} {\bibfield  {journal} {\bibinfo  {journal} {Reviews of Modern Physics}\ }\textbf {\bibinfo {volume} {91}},\ \bibinfo {pages} {035002} (\bibinfo {year} {2019})}\BibitemShut {NoStop}%
\bibitem [{\citenamefont {Scheibner}\ \emph {et~al.}(2022)\citenamefont {Scheibner}, \citenamefont {Fruchart},\ and\ \citenamefont {Vitelli}}]{Scheibner2022}%
  \BibitemOpen
  \bibfield  {author} {\bibinfo {author} {\bibfnamefont {C.}~\bibnamefont {Scheibner}}, \bibinfo {author} {\bibfnamefont {M.}~\bibnamefont {Fruchart}},\ and\ \bibinfo {author} {\bibfnamefont {V.}~\bibnamefont {Vitelli}},\ }\href@noop {} {\bibinfo {title} {Soft metamaterials: Adaptation and intelligence}} (\bibinfo {year} {2022}),\ \Eprint {https://arxiv.org/abs/2205.01867} {arXiv:2205.01867} \BibitemShut {NoStop}%
\bibitem [{\citenamefont {Murugan}\ \emph {et~al.}(2015)\citenamefont {Murugan}, \citenamefont {Zeravcic}, \citenamefont {Brenner},\ and\ \citenamefont {Leibler}}]{murugan_multifarious_2015}%
  \BibitemOpen
  \bibfield  {author} {\bibinfo {author} {\bibfnamefont {A.}~\bibnamefont {Murugan}}, \bibinfo {author} {\bibfnamefont {Z.}~\bibnamefont {Zeravcic}}, \bibinfo {author} {\bibfnamefont {M.~P.}\ \bibnamefont {Brenner}},\ and\ \bibinfo {author} {\bibfnamefont {S.}~\bibnamefont {Leibler}},\ }\bibfield  {title} {\bibinfo {title} {Multifarious assembly mixtures: {{Systems}} allowing retrieval of diverse stored structures},\ }\href {https://doi.org/10.1073/pnas.1413941112} {\bibfield  {journal} {\bibinfo  {journal} {Proceedings of the National Academy of Sciences}\ }\textbf {\bibinfo {volume} {112}},\ \bibinfo {pages} {54} (\bibinfo {year} {2015})}\BibitemShut {NoStop}%
\bibitem [{\citenamefont {Falk}\ \emph {et~al.}(2023{\natexlab{a}})\citenamefont {Falk}, \citenamefont {Wu}, \citenamefont {Matthews}, \citenamefont {Sachdeva}, \citenamefont {Pashine}, \citenamefont {Gardel}, \citenamefont {Nagel},\ and\ \citenamefont {Murugan}}]{Falk2023b}%
  \BibitemOpen
  \bibfield  {author} {\bibinfo {author} {\bibfnamefont {M.~J.}\ \bibnamefont {Falk}}, \bibinfo {author} {\bibfnamefont {J.}~\bibnamefont {Wu}}, \bibinfo {author} {\bibfnamefont {A.}~\bibnamefont {Matthews}}, \bibinfo {author} {\bibfnamefont {V.}~\bibnamefont {Sachdeva}}, \bibinfo {author} {\bibfnamefont {N.}~\bibnamefont {Pashine}}, \bibinfo {author} {\bibfnamefont {M.~L.}\ \bibnamefont {Gardel}}, \bibinfo {author} {\bibfnamefont {S.~R.}\ \bibnamefont {Nagel}},\ and\ \bibinfo {author} {\bibfnamefont {A.}~\bibnamefont {Murugan}},\ }\bibfield  {title} {\bibinfo {title} {Learning to learn by using nonequilibrium training protocols for adaptable materials},\ }\bibfield  {journal} {\bibinfo  {journal} {Proceedings of the National Academy of Sciences}\ }\textbf {\bibinfo {volume} {120}},\ \href {https://doi.org/10.1073/pnas.2219558120} {10.1073/pnas.2219558120} (\bibinfo {year} {2023}{\natexlab{a}})\BibitemShut {NoStop}%
\bibitem [{\citenamefont {Pashine}\ \emph {et~al.}(2019)\citenamefont {Pashine}, \citenamefont {Hexner}, \citenamefont {Liu},\ and\ \citenamefont {Nagel}}]{Pashine2019}%
  \BibitemOpen
  \bibfield  {author} {\bibinfo {author} {\bibfnamefont {N.}~\bibnamefont {Pashine}}, \bibinfo {author} {\bibfnamefont {D.}~\bibnamefont {Hexner}}, \bibinfo {author} {\bibfnamefont {A.~J.}\ \bibnamefont {Liu}},\ and\ \bibinfo {author} {\bibfnamefont {S.~R.}\ \bibnamefont {Nagel}},\ }\bibfield  {title} {\bibinfo {title} {Directed aging, memory, and nature’s greed},\ }\bibfield  {journal} {\bibinfo  {journal} {Science Advances}\ }\textbf {\bibinfo {volume} {5}},\ \href {https://doi.org/10.1126/sciadv.aax4215} {10.1126/sciadv.aax4215} (\bibinfo {year} {2019})\BibitemShut {NoStop}%
\bibitem [{\citenamefont {Stern}\ \emph {et~al.}(2022)\citenamefont {Stern}, \citenamefont {Dillavou}, \citenamefont {Miskin}, \citenamefont {Durian},\ and\ \citenamefont {Liu}}]{Stern2022}%
  \BibitemOpen
  \bibfield  {author} {\bibinfo {author} {\bibfnamefont {M.}~\bibnamefont {Stern}}, \bibinfo {author} {\bibfnamefont {S.}~\bibnamefont {Dillavou}}, \bibinfo {author} {\bibfnamefont {M.~Z.}\ \bibnamefont {Miskin}}, \bibinfo {author} {\bibfnamefont {D.~J.}\ \bibnamefont {Durian}},\ and\ \bibinfo {author} {\bibfnamefont {A.~J.}\ \bibnamefont {Liu}},\ }\bibfield  {title} {\bibinfo {title} {Physical learning beyond the quasistatic limit},\ }\href {https://doi.org/10.1103/physrevresearch.4.l022037} {\bibfield  {journal} {\bibinfo  {journal} {Physical Review Research}\ }\textbf {\bibinfo {volume} {4}},\ \bibinfo {pages} {l022037} (\bibinfo {year} {2022})}\BibitemShut {NoStop}%
\bibitem [{\citenamefont {McMullen}\ \emph {et~al.}(2022)\citenamefont {McMullen}, \citenamefont {Muñoz~Basagoiti}, \citenamefont {Zeravcic},\ and\ \citenamefont {Brujic}}]{McMullen2022}%
  \BibitemOpen
  \bibfield  {author} {\bibinfo {author} {\bibfnamefont {A.}~\bibnamefont {McMullen}}, \bibinfo {author} {\bibfnamefont {M.}~\bibnamefont {Muñoz~Basagoiti}}, \bibinfo {author} {\bibfnamefont {Z.}~\bibnamefont {Zeravcic}},\ and\ \bibinfo {author} {\bibfnamefont {J.}~\bibnamefont {Brujic}},\ }\bibfield  {title} {\bibinfo {title} {Self-assembly of emulsion droplets through programmable folding},\ }\href {https://doi.org/10.1038/s41586-022-05198-8} {\bibfield  {journal} {\bibinfo  {journal} {Nature}\ }\textbf {\bibinfo {volume} {610}},\ \bibinfo {pages} {502–506} (\bibinfo {year} {2022})}\BibitemShut {NoStop}%
\bibitem [{\citenamefont {Glotzer}(2004)}]{Glotzer2004}%
  \BibitemOpen
  \bibfield  {author} {\bibinfo {author} {\bibfnamefont {S.~C.}\ \bibnamefont {Glotzer}},\ }\bibfield  {title} {\bibinfo {title} {Some assembly required},\ }\href {https://doi.org/10.1126/science.1099988} {\bibfield  {journal} {\bibinfo  {journal} {Science}\ }\textbf {\bibinfo {volume} {306}},\ \bibinfo {pages} {419–420} (\bibinfo {year} {2004})}\BibitemShut {NoStop}%
\bibitem [{\citenamefont {Lee}\ \emph {et~al.}(2022)\citenamefont {Lee}, \citenamefont {Calcaterra}, \citenamefont {Lee}, \citenamefont {Hadibrata}, \citenamefont {Lee}, \citenamefont {Oh}, \citenamefont {Aydin}, \citenamefont {Glotzer},\ and\ \citenamefont {Mirkin}}]{Lee2022}%
  \BibitemOpen
  \bibfield  {author} {\bibinfo {author} {\bibfnamefont {S.}~\bibnamefont {Lee}}, \bibinfo {author} {\bibfnamefont {H.~A.}\ \bibnamefont {Calcaterra}}, \bibinfo {author} {\bibfnamefont {S.}~\bibnamefont {Lee}}, \bibinfo {author} {\bibfnamefont {W.}~\bibnamefont {Hadibrata}}, \bibinfo {author} {\bibfnamefont {B.}~\bibnamefont {Lee}}, \bibinfo {author} {\bibfnamefont {E.}~\bibnamefont {Oh}}, \bibinfo {author} {\bibfnamefont {K.}~\bibnamefont {Aydin}}, \bibinfo {author} {\bibfnamefont {S.~C.}\ \bibnamefont {Glotzer}},\ and\ \bibinfo {author} {\bibfnamefont {C.~A.}\ \bibnamefont {Mirkin}},\ }\bibfield  {title} {\bibinfo {title} {Shape memory in self-adapting colloidal crystals},\ }\href {https://doi.org/10.1038/s41586-022-05232-9} {\bibfield  {journal} {\bibinfo  {journal} {Nature}\ }\textbf {\bibinfo {volume} {610}},\ \bibinfo {pages} {674–679} (\bibinfo {year} {2022})}\BibitemShut {NoStop}%
\bibitem [{\citenamefont {Djellouli}\ \emph {et~al.}(2024)\citenamefont {Djellouli}, \citenamefont {Van~Raemdonck}, \citenamefont {Wang}, \citenamefont {Yang}, \citenamefont {Caillaud}, \citenamefont {Weitz}, \citenamefont {Rubinstein}, \citenamefont {Gorissen},\ and\ \citenamefont {Bertoldi}}]{Djellouli2024}%
  \BibitemOpen
  \bibfield  {author} {\bibinfo {author} {\bibfnamefont {A.}~\bibnamefont {Djellouli}}, \bibinfo {author} {\bibfnamefont {B.}~\bibnamefont {Van~Raemdonck}}, \bibinfo {author} {\bibfnamefont {Y.}~\bibnamefont {Wang}}, \bibinfo {author} {\bibfnamefont {Y.}~\bibnamefont {Yang}}, \bibinfo {author} {\bibfnamefont {A.}~\bibnamefont {Caillaud}}, \bibinfo {author} {\bibfnamefont {D.}~\bibnamefont {Weitz}}, \bibinfo {author} {\bibfnamefont {S.}~\bibnamefont {Rubinstein}}, \bibinfo {author} {\bibfnamefont {B.}~\bibnamefont {Gorissen}},\ and\ \bibinfo {author} {\bibfnamefont {K.}~\bibnamefont {Bertoldi}},\ }\bibfield  {title} {\bibinfo {title} {Shell buckling for programmable metafluids},\ }\href {https://doi.org/10.1038/s41586-024-07163-z} {\bibfield  {journal} {\bibinfo  {journal} {Nature}\ }\textbf {\bibinfo {volume} {628}},\ \bibinfo {pages} {545–550} (\bibinfo {year} {2024})}\BibitemShut {NoStop}%
\bibitem [{\citenamefont {Evans}\ \emph {et~al.}(2024)\citenamefont {Evans}, \citenamefont {O’Brien}, \citenamefont {Winfree},\ and\ \citenamefont {Murugan}}]{Evans2024}%
  \BibitemOpen
  \bibfield  {author} {\bibinfo {author} {\bibfnamefont {C.~G.}\ \bibnamefont {Evans}}, \bibinfo {author} {\bibfnamefont {J.}~\bibnamefont {O’Brien}}, \bibinfo {author} {\bibfnamefont {E.}~\bibnamefont {Winfree}},\ and\ \bibinfo {author} {\bibfnamefont {A.}~\bibnamefont {Murugan}},\ }\bibfield  {title} {\bibinfo {title} {Pattern recognition in the nucleation kinetics of non-equilibrium self-assembly},\ }\href {https://doi.org/10.1038/s41586-023-06890-z} {\bibfield  {journal} {\bibinfo  {journal} {Nature}\ }\textbf {\bibinfo {volume} {625}},\ \bibinfo {pages} {500–507} (\bibinfo {year} {2024})}\BibitemShut {NoStop}%
\bibitem [{\citenamefont {Wang}\ \emph {et~al.}(2023)\citenamefont {Wang}, \citenamefont {Pierce}, \citenamefont {Kojouharov}, \citenamefont {Chong}, \citenamefont {Diaz}, \citenamefont {Lu},\ and\ \citenamefont {Goldman}}]{Wang2023}%
  \BibitemOpen
  \bibfield  {author} {\bibinfo {author} {\bibfnamefont {T.}~\bibnamefont {Wang}}, \bibinfo {author} {\bibfnamefont {C.}~\bibnamefont {Pierce}}, \bibinfo {author} {\bibfnamefont {V.}~\bibnamefont {Kojouharov}}, \bibinfo {author} {\bibfnamefont {B.}~\bibnamefont {Chong}}, \bibinfo {author} {\bibfnamefont {K.}~\bibnamefont {Diaz}}, \bibinfo {author} {\bibfnamefont {H.}~\bibnamefont {Lu}},\ and\ \bibinfo {author} {\bibfnamefont {D.~I.}\ \bibnamefont {Goldman}},\ }\bibfield  {title} {\bibinfo {title} {Mechanical intelligence simplifies control in terrestrial limbless locomotion},\ }\bibfield  {journal} {\bibinfo  {journal} {Science Robotics}\ }\textbf {\bibinfo {volume} {8}},\ \href {https://doi.org/10.1126/scirobotics.adi2243} {10.1126/scirobotics.adi2243} (\bibinfo {year} {2023})\BibitemShut {NoStop}%
\bibitem [{\citenamefont {Aguilar}\ \emph {et~al.}(2016)\citenamefont {Aguilar}, \citenamefont {Zhang}, \citenamefont {Qian}, \citenamefont {Kingsbury}, \citenamefont {McInroe}, \citenamefont {Mazouchova}, \citenamefont {Li}, \citenamefont {Maladen}, \citenamefont {Gong}, \citenamefont {Travers}, \citenamefont {Hatton}, \citenamefont {Choset}, \citenamefont {Umbanhowar},\ and\ \citenamefont {Goldman}}]{Aguilar2016}%
  \BibitemOpen
  \bibfield  {author} {\bibinfo {author} {\bibfnamefont {J.}~\bibnamefont {Aguilar}}, \bibinfo {author} {\bibfnamefont {T.}~\bibnamefont {Zhang}}, \bibinfo {author} {\bibfnamefont {F.}~\bibnamefont {Qian}}, \bibinfo {author} {\bibfnamefont {M.}~\bibnamefont {Kingsbury}}, \bibinfo {author} {\bibfnamefont {B.}~\bibnamefont {McInroe}}, \bibinfo {author} {\bibfnamefont {N.}~\bibnamefont {Mazouchova}}, \bibinfo {author} {\bibfnamefont {C.}~\bibnamefont {Li}}, \bibinfo {author} {\bibfnamefont {R.}~\bibnamefont {Maladen}}, \bibinfo {author} {\bibfnamefont {C.}~\bibnamefont {Gong}}, \bibinfo {author} {\bibfnamefont {M.}~\bibnamefont {Travers}}, \bibinfo {author} {\bibfnamefont {R.~L.}\ \bibnamefont {Hatton}}, \bibinfo {author} {\bibfnamefont {H.}~\bibnamefont {Choset}}, \bibinfo {author} {\bibfnamefont {P.~B.}\ \bibnamefont {Umbanhowar}},\ and\ \bibinfo {author} {\bibfnamefont {D.~I.}\ \bibnamefont {Goldman}},\ }\bibfield  {title} {\bibinfo {title} {A review on locomotion robophysics: the study of movement at the
  intersection of robotics, soft matter and dynamical systems},\ }\href {https://doi.org/10.1088/0034-4885/79/11/110001} {\bibfield  {journal} {\bibinfo  {journal} {Reports on Progress in Physics}\ }\textbf {\bibinfo {volume} {79}},\ \bibinfo {pages} {110001} (\bibinfo {year} {2016})}\BibitemShut {NoStop}%
\bibitem [{\citenamefont {King}\ \emph {et~al.}(2023)\citenamefont {King}, \citenamefont {Du}, \citenamefont {Zhu}, \citenamefont {Schoenholz},\ and\ \citenamefont {Brenner}}]{King2023}%
  \BibitemOpen
  \bibfield  {author} {\bibinfo {author} {\bibfnamefont {E.~M.}\ \bibnamefont {King}}, \bibinfo {author} {\bibfnamefont {C.~X.}\ \bibnamefont {Du}}, \bibinfo {author} {\bibfnamefont {Q.-Z.}\ \bibnamefont {Zhu}}, \bibinfo {author} {\bibfnamefont {S.~S.}\ \bibnamefont {Schoenholz}},\ and\ \bibinfo {author} {\bibfnamefont {M.~P.}\ \bibnamefont {Brenner}},\ }\href@noop {} {\bibinfo {title} {Programmable patchy particles for materials design}} (\bibinfo {year} {2023}),\ \Eprint {https://arxiv.org/abs/2312.05360} {arXiv:2312.05360} \BibitemShut {NoStop}%
\bibitem [{\citenamefont {Niu}\ \emph {et~al.}(2019)\citenamefont {Niu}, \citenamefont {Du}, \citenamefont {Esposito}, \citenamefont {Ng}, \citenamefont {Brenner}, \citenamefont {McEuen},\ and\ \citenamefont {Cohen}}]{Niu2019}%
  \BibitemOpen
  \bibfield  {author} {\bibinfo {author} {\bibfnamefont {R.}~\bibnamefont {Niu}}, \bibinfo {author} {\bibfnamefont {C.~X.}\ \bibnamefont {Du}}, \bibinfo {author} {\bibfnamefont {E.}~\bibnamefont {Esposito}}, \bibinfo {author} {\bibfnamefont {J.}~\bibnamefont {Ng}}, \bibinfo {author} {\bibfnamefont {M.~P.}\ \bibnamefont {Brenner}}, \bibinfo {author} {\bibfnamefont {P.~L.}\ \bibnamefont {McEuen}},\ and\ \bibinfo {author} {\bibfnamefont {I.}~\bibnamefont {Cohen}},\ }\bibfield  {title} {\bibinfo {title} {Magnetic handshake materials as a scale-invariant platform for programmed self-assembly},\ }\href {https://doi.org/10.1073/pnas.1910332116} {\bibfield  {journal} {\bibinfo  {journal} {Proceedings of the National Academy of Sciences}\ }\textbf {\bibinfo {volume} {116}},\ \bibinfo {pages} {24402–24407} (\bibinfo {year} {2019})}\BibitemShut {NoStop}%
\bibitem [{\citenamefont {Meredith}\ \emph {et~al.}(2020)\citenamefont {Meredith}, \citenamefont {Moerman}, \citenamefont {Groenewold}, \citenamefont {Chiu}, \citenamefont {Kegel}, \citenamefont {van Blaaderen},\ and\ \citenamefont {Zarzar}}]{Meredith2020}%
  \BibitemOpen
  \bibfield  {author} {\bibinfo {author} {\bibfnamefont {C.~H.}\ \bibnamefont {Meredith}}, \bibinfo {author} {\bibfnamefont {P.~G.}\ \bibnamefont {Moerman}}, \bibinfo {author} {\bibfnamefont {J.}~\bibnamefont {Groenewold}}, \bibinfo {author} {\bibfnamefont {Y.-J.}\ \bibnamefont {Chiu}}, \bibinfo {author} {\bibfnamefont {W.~K.}\ \bibnamefont {Kegel}}, \bibinfo {author} {\bibfnamefont {A.}~\bibnamefont {van Blaaderen}},\ and\ \bibinfo {author} {\bibfnamefont {L.~D.}\ \bibnamefont {Zarzar}},\ }\bibfield  {title} {\bibinfo {title} {Predator–prey interactions between droplets driven by non-reciprocal oil exchange},\ }\href {https://doi.org/10.1038/s41557-020-00575-0} {\bibfield  {journal} {\bibinfo  {journal} {Nature Chemistry}\ }\textbf {\bibinfo {volume} {12}},\ \bibinfo {pages} {1136–1142} (\bibinfo {year} {2020})}\BibitemShut {NoStop}%
\bibitem [{\citenamefont {Moran}\ and\ \citenamefont {Posner}(2017)}]{Moran2017}%
  \BibitemOpen
  \bibfield  {author} {\bibinfo {author} {\bibfnamefont {J.~L.}\ \bibnamefont {Moran}}\ and\ \bibinfo {author} {\bibfnamefont {J.~D.}\ \bibnamefont {Posner}},\ }\bibfield  {title} {\bibinfo {title} {Phoretic self-propulsion},\ }\href {https://doi.org/10.1146/annurev-fluid-122414-034456} {\bibfield  {journal} {\bibinfo  {journal} {Annual Review of Fluid Mechanics}\ }\textbf {\bibinfo {volume} {49}},\ \bibinfo {pages} {511–540} (\bibinfo {year} {2017})}\BibitemShut {NoStop}%
\bibitem [{\citenamefont {Michelin}(2023)}]{Michelin2023}%
  \BibitemOpen
  \bibfield  {author} {\bibinfo {author} {\bibfnamefont {S.}~\bibnamefont {Michelin}},\ }\bibfield  {title} {\bibinfo {title} {Self-propulsion of chemically active droplets},\ }\href {https://doi.org/10.1146/annurev-fluid-120720-012204} {\bibfield  {journal} {\bibinfo  {journal} {Annual Review of Fluid Mechanics}\ }\textbf {\bibinfo {volume} {55}},\ \bibinfo {pages} {77–101} (\bibinfo {year} {2023})}\BibitemShut {NoStop}%
\bibitem [{\citenamefont {Nguyen}\ and\ \citenamefont {Stebe}(2002)}]{Nguyen2002}%
  \BibitemOpen
  \bibfield  {author} {\bibinfo {author} {\bibfnamefont {V.~X.}\ \bibnamefont {Nguyen}}\ and\ \bibinfo {author} {\bibfnamefont {K.~J.}\ \bibnamefont {Stebe}},\ }\bibfield  {title} {\bibinfo {title} {Patterning of small particles by a surfactant-enhanced marangoni-bénard instability},\ }\href {https://doi.org/10.1103/physrevlett.88.164501} {\bibfield  {journal} {\bibinfo  {journal} {Physical Review Letters}\ }\textbf {\bibinfo {volume} {88}},\ \bibinfo {pages} {164501} (\bibinfo {year} {2002})}\BibitemShut {NoStop}%
\bibitem [{\citenamefont {Maass}\ \emph {et~al.}(2016)\citenamefont {Maass}, \citenamefont {Krüger}, \citenamefont {Herminghaus},\ and\ \citenamefont {Bahr}}]{Maass2016}%
  \BibitemOpen
  \bibfield  {author} {\bibinfo {author} {\bibfnamefont {C.~C.}\ \bibnamefont {Maass}}, \bibinfo {author} {\bibfnamefont {C.}~\bibnamefont {Krüger}}, \bibinfo {author} {\bibfnamefont {S.}~\bibnamefont {Herminghaus}},\ and\ \bibinfo {author} {\bibfnamefont {C.}~\bibnamefont {Bahr}},\ }\bibfield  {title} {\bibinfo {title} {Swimming droplets},\ }\href {https://doi.org/10.1146/annurev-conmatphys-031115-011517} {\bibfield  {journal} {\bibinfo  {journal} {Annual Review of Condensed Matter Physics}\ }\textbf {\bibinfo {volume} {7}},\ \bibinfo {pages} {171–193} (\bibinfo {year} {2016})}\BibitemShut {NoStop}%
\bibitem [{\citenamefont {Michelin}\ \emph {et~al.}(2013)\citenamefont {Michelin}, \citenamefont {Lauga},\ and\ \citenamefont {Bartolo}}]{Michelin2013}%
  \BibitemOpen
  \bibfield  {author} {\bibinfo {author} {\bibfnamefont {S.}~\bibnamefont {Michelin}}, \bibinfo {author} {\bibfnamefont {E.}~\bibnamefont {Lauga}},\ and\ \bibinfo {author} {\bibfnamefont {D.}~\bibnamefont {Bartolo}},\ }\bibfield  {title} {\bibinfo {title} {Spontaneous autophoretic motion of isotropic particles},\ }\bibfield  {journal} {\bibinfo  {journal} {Physics of Fluids}\ }\textbf {\bibinfo {volume} {25}},\ \href {https://doi.org/10.1063/1.4810749} {10.1063/1.4810749} (\bibinfo {year} {2013})\BibitemShut {NoStop}%
\bibitem [{\citenamefont {Izri}\ \emph {et~al.}(2014)\citenamefont {Izri}, \citenamefont {van~der Linden}, \citenamefont {Michelin},\ and\ \citenamefont {Dauchot}}]{Izri2014}%
  \BibitemOpen
  \bibfield  {author} {\bibinfo {author} {\bibfnamefont {Z.}~\bibnamefont {Izri}}, \bibinfo {author} {\bibfnamefont {M.~N.}\ \bibnamefont {van~der Linden}}, \bibinfo {author} {\bibfnamefont {S.}~\bibnamefont {Michelin}},\ and\ \bibinfo {author} {\bibfnamefont {O.}~\bibnamefont {Dauchot}},\ }\bibfield  {title} {\bibinfo {title} {Self-propulsion of pure water droplets by spontaneous marangoni-stress-driven motion},\ }\href {https://doi.org/10.1103/physrevlett.113.248302} {\bibfield  {journal} {\bibinfo  {journal} {Physical Review Letters}\ }\textbf {\bibinfo {volume} {113}},\ \bibinfo {pages} {248302} (\bibinfo {year} {2014})}\BibitemShut {NoStop}%
\bibitem [{\citenamefont {Schmitt}\ and\ \citenamefont {Stark}(2013)}]{Schmitt2013}%
  \BibitemOpen
  \bibfield  {author} {\bibinfo {author} {\bibfnamefont {M.}~\bibnamefont {Schmitt}}\ and\ \bibinfo {author} {\bibfnamefont {H.}~\bibnamefont {Stark}},\ }\bibfield  {title} {\bibinfo {title} {Swimming active droplet: A theoretical analysis},\ }\href {https://doi.org/10.1209/0295-5075/101/44008} {\bibfield  {journal} {\bibinfo  {journal} {EPL (Europhysics Letters)}\ }\textbf {\bibinfo {volume} {101}},\ \bibinfo {pages} {44008} (\bibinfo {year} {2013})}\BibitemShut {NoStop}%
\bibitem [{\citenamefont {Soto}\ and\ \citenamefont {Golestanian}(2014)}]{Soto2014}%
  \BibitemOpen
  \bibfield  {author} {\bibinfo {author} {\bibfnamefont {R.}~\bibnamefont {Soto}}\ and\ \bibinfo {author} {\bibfnamefont {R.}~\bibnamefont {Golestanian}},\ }\bibfield  {title} {\bibinfo {title} {Self-assembly of catalytically active colloidal molecules: Tailoring activity through surface chemistry},\ }\href {https://doi.org/10.1103/physrevlett.112.068301} {\bibfield  {journal} {\bibinfo  {journal} {Physical Review Letters}\ }\textbf {\bibinfo {volume} {112}},\ \bibinfo {pages} {068301} (\bibinfo {year} {2014})}\BibitemShut {NoStop}%
\bibitem [{\citenamefont {Moerman}\ \emph {et~al.}(2017)\citenamefont {Moerman}, \citenamefont {Moyses}, \citenamefont {van~der Wee}, \citenamefont {Grier}, \citenamefont {van Blaaderen}, \citenamefont {Kegel}, \citenamefont {Groenewold},\ and\ \citenamefont {Brujic}}]{Moerman2017}%
  \BibitemOpen
  \bibfield  {author} {\bibinfo {author} {\bibfnamefont {P.~G.}\ \bibnamefont {Moerman}}, \bibinfo {author} {\bibfnamefont {H.~W.}\ \bibnamefont {Moyses}}, \bibinfo {author} {\bibfnamefont {E.~B.}\ \bibnamefont {van~der Wee}}, \bibinfo {author} {\bibfnamefont {D.~G.}\ \bibnamefont {Grier}}, \bibinfo {author} {\bibfnamefont {A.}~\bibnamefont {van Blaaderen}}, \bibinfo {author} {\bibfnamefont {W.~K.}\ \bibnamefont {Kegel}}, \bibinfo {author} {\bibfnamefont {J.}~\bibnamefont {Groenewold}},\ and\ \bibinfo {author} {\bibfnamefont {J.}~\bibnamefont {Brujic}},\ }\bibfield  {title} {\bibinfo {title} {Solute-mediated interactions between active droplets},\ }\href {https://doi.org/10.1103/physreve.96.032607} {\bibfield  {journal} {\bibinfo  {journal} {Physical Review E}\ }\textbf {\bibinfo {volume} {96}},\ \bibinfo {pages} {032607} (\bibinfo {year} {2017})}\BibitemShut {NoStop}%
\bibitem [{\citenamefont {Parisi}(1986)}]{Parisi1986}%
  \BibitemOpen
  \bibfield  {author} {\bibinfo {author} {\bibfnamefont {G.}~\bibnamefont {Parisi}},\ }\href {https://doi.org/10.1088/0305-4470/19/11/005} {\bibfield  {journal} {\bibinfo  {journal} {Journal of Physics A: Mathematical and General}\ }\textbf {\bibinfo {volume} {19}},\ \bibinfo {pages} {L675–L680} (\bibinfo {year} {1986})}\BibitemShut {NoStop}%
\bibitem [{\citenamefont {Crisanti}\ and\ \citenamefont {Sompolinsky}(1987)}]{Crisanti1987}%
  \BibitemOpen
  \bibfield  {author} {\bibinfo {author} {\bibfnamefont {A.}~\bibnamefont {Crisanti}}\ and\ \bibinfo {author} {\bibfnamefont {H.}~\bibnamefont {Sompolinsky}},\ }\bibfield  {title} {\bibinfo {title} {Dynamics of spin systems with randomly asymmetric bonds: Langevin dynamics and a spherical model},\ }\href {https://doi.org/10.1103/physreva.36.4922} {\bibfield  {journal} {\bibinfo  {journal} {Physical Review A}\ }\textbf {\bibinfo {volume} {36}},\ \bibinfo {pages} {4922–4939} (\bibinfo {year} {1987})}\BibitemShut {NoStop}%
\bibitem [{\citenamefont {Sompolinsky}\ and\ \citenamefont {Kanter}(1986)}]{Sompolinsky1986}%
  \BibitemOpen
  \bibfield  {author} {\bibinfo {author} {\bibfnamefont {H.}~\bibnamefont {Sompolinsky}}\ and\ \bibinfo {author} {\bibfnamefont {I.}~\bibnamefont {Kanter}},\ }\bibfield  {title} {\bibinfo {title} {Temporal association in asymmetric neural networks},\ }\href {https://doi.org/10.1103/physrevlett.57.2861} {\bibfield  {journal} {\bibinfo  {journal} {Physical Review Letters}\ }\textbf {\bibinfo {volume} {57}},\ \bibinfo {pages} {2861–2864} (\bibinfo {year} {1986})}\BibitemShut {NoStop}%
\bibitem [{\citenamefont {Kanter}\ and\ \citenamefont {Sompolinsky}(1987)}]{Kanter1987}%
  \BibitemOpen
  \bibfield  {author} {\bibinfo {author} {\bibfnamefont {I.}~\bibnamefont {Kanter}}\ and\ \bibinfo {author} {\bibfnamefont {H.}~\bibnamefont {Sompolinsky}},\ }\bibfield  {title} {\bibinfo {title} {Mean-field theory of spin-glasses with finite coordination number},\ }\href {https://doi.org/10.1103/physrevlett.58.164} {\bibfield  {journal} {\bibinfo  {journal} {Physical Review Letters}\ }\textbf {\bibinfo {volume} {58}},\ \bibinfo {pages} {164–167} (\bibinfo {year} {1987})}\BibitemShut {NoStop}%
\bibitem [{\citenamefont {Kleinfeld}(1986)}]{Kleinfeld1986}%
  \BibitemOpen
  \bibfield  {author} {\bibinfo {author} {\bibfnamefont {D.}~\bibnamefont {Kleinfeld}},\ }\bibfield  {title} {\bibinfo {title} {Sequential state generation by model neural networks.},\ }\href {https://doi.org/10.1073/pnas.83.24.9469} {\bibfield  {journal} {\bibinfo  {journal} {Proceedings of the National Academy of Sciences}\ }\textbf {\bibinfo {volume} {83}},\ \bibinfo {pages} {9469–9473} (\bibinfo {year} {1986})}\BibitemShut {NoStop}%
\bibitem [{\citenamefont {Herron}\ \emph {et~al.}(2023)\citenamefont {Herron}, \citenamefont {Sartori},\ and\ \citenamefont {Xue}}]{Herron2023}%
  \BibitemOpen
  \bibfield  {author} {\bibinfo {author} {\bibfnamefont {L.}~\bibnamefont {Herron}}, \bibinfo {author} {\bibfnamefont {P.}~\bibnamefont {Sartori}},\ and\ \bibinfo {author} {\bibfnamefont {B.}~\bibnamefont {Xue}},\ }\bibfield  {title} {\bibinfo {title} {Robust retrieval of dynamic sequences through interaction modulation},\ }\href {https://doi.org/10.1103/prxlife.1.023012} {\bibfield  {journal} {\bibinfo  {journal} {PRX Life}\ }\textbf {\bibinfo {volume} {1}},\ \bibinfo {pages} {023012} (\bibinfo {year} {2023})}\BibitemShut {NoStop}%
\bibitem [{\citenamefont {Dehaene}\ \emph {et~al.}(1987)\citenamefont {Dehaene}, \citenamefont {Changeux},\ and\ \citenamefont {Nadal}}]{Dehaene1987}%
  \BibitemOpen
  \bibfield  {author} {\bibinfo {author} {\bibfnamefont {S.}~\bibnamefont {Dehaene}}, \bibinfo {author} {\bibfnamefont {J.~P.}\ \bibnamefont {Changeux}},\ and\ \bibinfo {author} {\bibfnamefont {J.~P.}\ \bibnamefont {Nadal}},\ }\bibfield  {title} {\bibinfo {title} {Neural networks that learn temporal sequences by selection.},\ }\href {https://doi.org/10.1073/pnas.84.9.2727} {\bibfield  {journal} {\bibinfo  {journal} {Proceedings of the National Academy of Sciences}\ }\textbf {\bibinfo {volume} {84}},\ \bibinfo {pages} {2727–2731} (\bibinfo {year} {1987})}\BibitemShut {NoStop}%
\bibitem [{\citenamefont {Buhmann}\ and\ \citenamefont {Schulten}(1987)}]{Buhmann1987}%
  \BibitemOpen
  \bibfield  {author} {\bibinfo {author} {\bibfnamefont {J.}~\bibnamefont {Buhmann}}\ and\ \bibinfo {author} {\bibfnamefont {K.}~\bibnamefont {Schulten}},\ }\bibfield  {title} {\bibinfo {title} {Noise-driven temporal association in neural networks},\ }\href {https://doi.org/10.1209/0295-5075/4/10/021} {\bibfield  {journal} {\bibinfo  {journal} {Europhysics Letters (EPL)}\ }\textbf {\bibinfo {volume} {4}},\ \bibinfo {pages} {1205–1209} (\bibinfo {year} {1987})}\BibitemShut {NoStop}%
\bibitem [{\citenamefont {Kleinfeld}\ and\ \citenamefont {Sompolinsky}(1988)}]{Kleinfeld1988}%
  \BibitemOpen
  \bibfield  {author} {\bibinfo {author} {\bibfnamefont {D.}~\bibnamefont {Kleinfeld}}\ and\ \bibinfo {author} {\bibfnamefont {H.}~\bibnamefont {Sompolinsky}},\ }\bibfield  {title} {\bibinfo {title} {Associative neural network model for the generation of temporal patterns. theory and application to central pattern generators},\ }\href {https://doi.org/10.1016/s0006-3495(88)83041-8} {\bibfield  {journal} {\bibinfo  {journal} {Biophysical Journal}\ }\textbf {\bibinfo {volume} {54}},\ \bibinfo {pages} {1039–1051} (\bibinfo {year} {1988})}\BibitemShut {NoStop}%
\bibitem [{\citenamefont {Hertz}\ \emph {et~al.}(1987)\citenamefont {Hertz}, \citenamefont {Grinstein},\ and\ \citenamefont {Solla}}]{Hertz1987}%
  \BibitemOpen
  \bibfield  {author} {\bibinfo {author} {\bibfnamefont {J.~A.}\ \bibnamefont {Hertz}}, \bibinfo {author} {\bibfnamefont {G.}~\bibnamefont {Grinstein}},\ and\ \bibinfo {author} {\bibfnamefont {S.~A.}\ \bibnamefont {Solla}},\ }\bibinfo {title} {Irreversible spin glasses and neural networks},\ in\ \href {https://doi.org/10.1007/bfb0057533} {\emph {\bibinfo {booktitle} {Lecture Notes in Physics}}}\ (\bibinfo  {publisher} {Springer Berlin Heidelberg},\ \bibinfo {year} {1987})\ p.\ \bibinfo {pages} {538–546}\BibitemShut {NoStop}%
\bibitem [{\citenamefont {Hertz}\ \emph {et~al.}(1986)\citenamefont {Hertz}, \citenamefont {Grinstein},\ and\ \citenamefont {Solla}}]{Hertz1986}%
  \BibitemOpen
  \bibfield  {author} {\bibinfo {author} {\bibfnamefont {J.~A.}\ \bibnamefont {Hertz}}, \bibinfo {author} {\bibfnamefont {G.}~\bibnamefont {Grinstein}},\ and\ \bibinfo {author} {\bibfnamefont {S.~A.}\ \bibnamefont {Solla}},\ }\bibfield  {title} {\bibinfo {title} {Memory networks with asymmetric bonds},\ }in\ \href {https://doi.org/10.1063/1.36259} {\emph {\bibinfo {booktitle} {AIP Conference Proceedings}}}\ (\bibinfo  {publisher} {AIP},\ \bibinfo {year} {1986})\BibitemShut {NoStop}%
\bibitem [{\citenamefont {Ivlev}\ \emph {et~al.}(2015)\citenamefont {Ivlev}, \citenamefont {Bartnick}, \citenamefont {Heinen}, \citenamefont {Du}, \citenamefont {Nosenko},\ and\ \citenamefont {L\"{o}wen}}]{Ivlev2015}%
  \BibitemOpen
  \bibfield  {author} {\bibinfo {author} {\bibfnamefont {A.~V.}\ \bibnamefont {Ivlev}}, \bibinfo {author} {\bibfnamefont {J.}~\bibnamefont {Bartnick}}, \bibinfo {author} {\bibfnamefont {M.}~\bibnamefont {Heinen}}, \bibinfo {author} {\bibfnamefont {C.-R.}\ \bibnamefont {Du}}, \bibinfo {author} {\bibfnamefont {V.}~\bibnamefont {Nosenko}},\ and\ \bibinfo {author} {\bibfnamefont {H.}~\bibnamefont {L\"{o}wen}},\ }\bibfield  {title} {\bibinfo {title} {Statistical mechanics where newton’s third law is broken},\ }\href {https://doi.org/10.1103/physrevx.5.011035} {\bibfield  {journal} {\bibinfo  {journal} {Physical Review X}\ }\textbf {\bibinfo {volume} {5}},\ \bibinfo {pages} {011035} (\bibinfo {year} {2015})}\BibitemShut {NoStop}%
\bibitem [{\citenamefont {Fruchart}\ \emph {et~al.}(2021)\citenamefont {Fruchart}, \citenamefont {Hanai}, \citenamefont {Littlewood},\ and\ \citenamefont {Vitelli}}]{Fruchart2021}%
  \BibitemOpen
  \bibfield  {author} {\bibinfo {author} {\bibfnamefont {M.}~\bibnamefont {Fruchart}}, \bibinfo {author} {\bibfnamefont {R.}~\bibnamefont {Hanai}}, \bibinfo {author} {\bibfnamefont {P.~B.}\ \bibnamefont {Littlewood}},\ and\ \bibinfo {author} {\bibfnamefont {V.}~\bibnamefont {Vitelli}},\ }\bibfield  {title} {\bibinfo {title} {Non-reciprocal phase transitions},\ }\href@noop {} {\bibfield  {journal} {\bibinfo  {journal} {Nature}\ }\textbf {\bibinfo {volume} {592}},\ \bibinfo {pages} {363} (\bibinfo {year} {2021})}\BibitemShut {NoStop}%
\bibitem [{\citenamefont {Saha}\ \emph {et~al.}(2020)\citenamefont {Saha}, \citenamefont {Agudo-Canalejo},\ and\ \citenamefont {Golestanian}}]{Saha2020}%
  \BibitemOpen
  \bibfield  {author} {\bibinfo {author} {\bibfnamefont {S.}~\bibnamefont {Saha}}, \bibinfo {author} {\bibfnamefont {J.}~\bibnamefont {Agudo-Canalejo}},\ and\ \bibinfo {author} {\bibfnamefont {R.}~\bibnamefont {Golestanian}},\ }\bibfield  {title} {\bibinfo {title} {Scalar active mixtures: The nonreciprocal cahn-hilliard model},\ }\href@noop {} {\bibfield  {journal} {\bibinfo  {journal} {Physical Review X}\ }\textbf {\bibinfo {volume} {10}},\ \bibinfo {pages} {041009} (\bibinfo {year} {2020})}\BibitemShut {NoStop}%
\bibitem [{\citenamefont {You}\ \emph {et~al.}(2020)\citenamefont {You}, \citenamefont {Baskaran},\ and\ \citenamefont {Marchetti}}]{You2020}%
  \BibitemOpen
  \bibfield  {author} {\bibinfo {author} {\bibfnamefont {Z.}~\bibnamefont {You}}, \bibinfo {author} {\bibfnamefont {A.}~\bibnamefont {Baskaran}},\ and\ \bibinfo {author} {\bibfnamefont {M.~C.}\ \bibnamefont {Marchetti}},\ }\bibfield  {title} {\bibinfo {title} {Nonreciprocity as a generic route to traveling states},\ }\href@noop {} {\bibfield  {journal} {\bibinfo  {journal} {Proceedings of the National Academy of Sciences}\ }\textbf {\bibinfo {volume} {117}},\ \bibinfo {pages} {19767} (\bibinfo {year} {2020})}\BibitemShut {NoStop}%
\bibitem [{\citenamefont {Yifat}\ \emph {et~al.}(2018)\citenamefont {Yifat}, \citenamefont {Coursault}, \citenamefont {Peterson}, \citenamefont {Parker}, \citenamefont {Bao}, \citenamefont {Gray}, \citenamefont {Rice},\ and\ \citenamefont {Scherer}}]{Yifat2018}%
  \BibitemOpen
  \bibfield  {author} {\bibinfo {author} {\bibfnamefont {Y.}~\bibnamefont {Yifat}}, \bibinfo {author} {\bibfnamefont {D.}~\bibnamefont {Coursault}}, \bibinfo {author} {\bibfnamefont {C.~W.}\ \bibnamefont {Peterson}}, \bibinfo {author} {\bibfnamefont {J.}~\bibnamefont {Parker}}, \bibinfo {author} {\bibfnamefont {Y.}~\bibnamefont {Bao}}, \bibinfo {author} {\bibfnamefont {S.~K.}\ \bibnamefont {Gray}}, \bibinfo {author} {\bibfnamefont {S.~A.}\ \bibnamefont {Rice}},\ and\ \bibinfo {author} {\bibfnamefont {N.~F.}\ \bibnamefont {Scherer}},\ }\bibfield  {title} {\bibinfo {title} {Reactive optical matter: light-induced motility in electrodynamically asymmetric nanoscale scatterers},\ }\bibfield  {journal} {\bibinfo  {journal} {Light: Science and Applications}\ }\textbf {\bibinfo {volume} {7}},\ \href {https://doi.org/10.1038/s41377-018-0105-y} {10.1038/s41377-018-0105-y} (\bibinfo {year} {2018})\BibitemShut {NoStop}%
\bibitem [{\citenamefont {Drescher}\ \emph {et~al.}(2009)\citenamefont {Drescher}, \citenamefont {Leptos}, \citenamefont {Tuval}, \citenamefont {Ishikawa}, \citenamefont {Pedley},\ and\ \citenamefont {Goldstein}}]{Drescher2009}%
  \BibitemOpen
  \bibfield  {author} {\bibinfo {author} {\bibfnamefont {K.}~\bibnamefont {Drescher}}, \bibinfo {author} {\bibfnamefont {K.~C.}\ \bibnamefont {Leptos}}, \bibinfo {author} {\bibfnamefont {I.}~\bibnamefont {Tuval}}, \bibinfo {author} {\bibfnamefont {T.}~\bibnamefont {Ishikawa}}, \bibinfo {author} {\bibfnamefont {T.~J.}\ \bibnamefont {Pedley}},\ and\ \bibinfo {author} {\bibfnamefont {R.~E.}\ \bibnamefont {Goldstein}},\ }\bibfield  {title} {\bibinfo {title} {Dancing volvox: Hydrodynamic bound states of swimming algae},\ }\href {https://doi.org/10.1103/physrevlett.102.168101} {\bibfield  {journal} {\bibinfo  {journal} {Physical Review Letters}\ }\textbf {\bibinfo {volume} {102}},\ \bibinfo {pages} {168101} (\bibinfo {year} {2009})}\BibitemShut {NoStop}%
\bibitem [{\citenamefont {Osat}\ and\ \citenamefont {Golestanian}(2022)}]{Osat2022}%
  \BibitemOpen
  \bibfield  {author} {\bibinfo {author} {\bibfnamefont {S.}~\bibnamefont {Osat}}\ and\ \bibinfo {author} {\bibfnamefont {R.}~\bibnamefont {Golestanian}},\ }\bibfield  {title} {\bibinfo {title} {Non-reciprocal multifarious self-organization},\ }\href {https://doi.org/10.1038/s41565-022-01258-2} {\bibfield  {journal} {\bibinfo  {journal} {Nature Nanotechnology}\ }\textbf {\bibinfo {volume} {18}},\ \bibinfo {pages} {79–85} (\bibinfo {year} {2022})}\BibitemShut {NoStop}%
\bibitem [{\citenamefont {Baek}\ \emph {et~al.}(2018)\citenamefont {Baek}, \citenamefont {Solon}, \citenamefont {Xu}, \citenamefont {Nikola},\ and\ \citenamefont {Kafri}}]{Baek2018}%
  \BibitemOpen
  \bibfield  {author} {\bibinfo {author} {\bibfnamefont {Y.}~\bibnamefont {Baek}}, \bibinfo {author} {\bibfnamefont {A.~P.}\ \bibnamefont {Solon}}, \bibinfo {author} {\bibfnamefont {X.}~\bibnamefont {Xu}}, \bibinfo {author} {\bibfnamefont {N.}~\bibnamefont {Nikola}},\ and\ \bibinfo {author} {\bibfnamefont {Y.}~\bibnamefont {Kafri}},\ }\bibfield  {title} {\bibinfo {title} {Generic long-range interactions between passive bodies in an active fluid},\ }\href {https://doi.org/10.1103/physrevlett.120.058002} {\bibfield  {journal} {\bibinfo  {journal} {Physical Review Letters}\ }\textbf {\bibinfo {volume} {120}},\ \bibinfo {pages} {058002} (\bibinfo {year} {2018})}\BibitemShut {NoStop}%
\bibitem [{\citenamefont {Banerjee}\ \emph {et~al.}(2022)\citenamefont {Banerjee}, \citenamefont {Mandal}, \citenamefont {Banerjee}, \citenamefont {Thutupalli},\ and\ \citenamefont {Rao}}]{banerjee2022}%
  \BibitemOpen
  \bibfield  {author} {\bibinfo {author} {\bibfnamefont {J.~P.}\ \bibnamefont {Banerjee}}, \bibinfo {author} {\bibfnamefont {R.}~\bibnamefont {Mandal}}, \bibinfo {author} {\bibfnamefont {D.~S.}\ \bibnamefont {Banerjee}}, \bibinfo {author} {\bibfnamefont {S.}~\bibnamefont {Thutupalli}},\ and\ \bibinfo {author} {\bibfnamefont {M.}~\bibnamefont {Rao}},\ }\bibfield  {title} {\bibinfo {title} {Unjamming and emergent nonreciprocity in active ploughing through a compressible viscoelastic fluid},\ }\href {https://doi.org/10.1038/s41467-022-31984-z} {\bibfield  {journal} {\bibinfo  {journal} {Nature Communications}\ }\textbf {\bibinfo {volume} {13}},\ \bibinfo {pages} {4533} (\bibinfo {year} {2022})}\BibitemShut {NoStop}%
\bibitem [{\citenamefont {Dinelli}\ \emph {et~al.}(2023)\citenamefont {Dinelli}, \citenamefont {O’Byrne}, \citenamefont {Curatolo}, \citenamefont {Zhao}, \citenamefont {Sollich},\ and\ \citenamefont {Tailleur}}]{Dinelli2023}%
  \BibitemOpen
  \bibfield  {author} {\bibinfo {author} {\bibfnamefont {A.}~\bibnamefont {Dinelli}}, \bibinfo {author} {\bibfnamefont {J.}~\bibnamefont {O’Byrne}}, \bibinfo {author} {\bibfnamefont {A.}~\bibnamefont {Curatolo}}, \bibinfo {author} {\bibfnamefont {Y.}~\bibnamefont {Zhao}}, \bibinfo {author} {\bibfnamefont {P.}~\bibnamefont {Sollich}},\ and\ \bibinfo {author} {\bibfnamefont {J.}~\bibnamefont {Tailleur}},\ }\bibfield  {title} {\bibinfo {title} {Non-reciprocity across scales in active mixtures},\ }\bibfield  {journal} {\bibinfo  {journal} {Nature Communications}\ }\textbf {\bibinfo {volume} {14}},\ \href {https://doi.org/10.1038/s41467-023-42713-5} {10.1038/s41467-023-42713-5} (\bibinfo {year} {2023})\BibitemShut {NoStop}%
\bibitem [{\citenamefont {Mandal}\ \emph {et~al.}(2022)\citenamefont {Mandal}, \citenamefont {Jaramillo},\ and\ \citenamefont {Sollich}}]{mandal2022}%
  \BibitemOpen
  \bibfield  {author} {\bibinfo {author} {\bibfnamefont {R.}~\bibnamefont {Mandal}}, \bibinfo {author} {\bibfnamefont {S.~S.}\ \bibnamefont {Jaramillo}},\ and\ \bibinfo {author} {\bibfnamefont {P.}~\bibnamefont {Sollich}},\ }\href@noop {} {\bibinfo {title} {Robustness of travelling states in generic non-reciprocal mixtures}} (\bibinfo {year} {2022}),\ \Eprint {https://arxiv.org/abs/2212.05618} {arXiv:2212.05618 [cond-mat.stat-mech]} \BibitemShut {NoStop}%
\bibitem [{\citenamefont {Avni}\ \emph {et~al.}(2023)\citenamefont {Avni}, \citenamefont {Fruchart}, \citenamefont {Martin}, \citenamefont {Seara},\ and\ \citenamefont {Vitelli}}]{Avni2023}%
  \BibitemOpen
  \bibfield  {author} {\bibinfo {author} {\bibfnamefont {Y.}~\bibnamefont {Avni}}, \bibinfo {author} {\bibfnamefont {M.}~\bibnamefont {Fruchart}}, \bibinfo {author} {\bibfnamefont {D.}~\bibnamefont {Martin}}, \bibinfo {author} {\bibfnamefont {D.}~\bibnamefont {Seara}},\ and\ \bibinfo {author} {\bibfnamefont {V.}~\bibnamefont {Vitelli}},\ }\href@noop {} {\bibinfo {title} {The non-reciprocal ising model}} (\bibinfo {year} {2023}),\ \Eprint {https://arxiv.org/abs/2311.05471} {arXiv:2311.05471} \BibitemShut {NoStop}%
\bibitem [{\citenamefont {Abbott}\ and\ \citenamefont {Nelson}(2000)}]{Abbott2000}%
  \BibitemOpen
  \bibfield  {author} {\bibinfo {author} {\bibfnamefont {L.~F.}\ \bibnamefont {Abbott}}\ and\ \bibinfo {author} {\bibfnamefont {S.~B.}\ \bibnamefont {Nelson}},\ }\bibfield  {title} {\bibinfo {title} {Synaptic plasticity: taming the beast},\ }\href {https://doi.org/10.1038/81453} {\bibfield  {journal} {\bibinfo  {journal} {Nature Neuroscience}\ }\textbf {\bibinfo {volume} {3}},\ \bibinfo {pages} {1178–1183} (\bibinfo {year} {2000})}\BibitemShut {NoStop}%
\bibitem [{\citenamefont {Dan}\ and\ \citenamefont {Poo}(2004)}]{Dan2004}%
  \BibitemOpen
  \bibfield  {author} {\bibinfo {author} {\bibfnamefont {Y.}~\bibnamefont {Dan}}\ and\ \bibinfo {author} {\bibfnamefont {M.-m.}\ \bibnamefont {Poo}},\ }\bibfield  {title} {\bibinfo {title} {Spike timing-dependent plasticity of neural circuits},\ }\href {https://doi.org/10.1016/j.neuron.2004.09.007} {\bibfield  {journal} {\bibinfo  {journal} {Neuron}\ }\textbf {\bibinfo {volume} {44}},\ \bibinfo {pages} {23–30} (\bibinfo {year} {2004})}\BibitemShut {NoStop}%
\bibitem [{\citenamefont {Caporale}\ and\ \citenamefont {Dan}(2008)}]{caporale_spike_2008}%
  \BibitemOpen
  \bibfield  {author} {\bibinfo {author} {\bibfnamefont {N.}~\bibnamefont {Caporale}}\ and\ \bibinfo {author} {\bibfnamefont {Y.}~\bibnamefont {Dan}},\ }\bibfield  {title} {\bibinfo {title} {Spike {{Timing}}{\textendash}{{Dependent Plasticity}}: {{A Hebbian Learning Rule}}},\ }\href {https://doi.org/10.1146/annurev.neuro.31.060407.125639} {\bibfield  {journal} {\bibinfo  {journal} {Annual Review of Neuroscience}\ }\textbf {\bibinfo {volume} {31}},\ \bibinfo {pages} {25} (\bibinfo {year} {2008})}\BibitemShut {NoStop}%
\bibitem [{\citenamefont {Shouval}(2010)}]{shouval_spike_2010}%
  \BibitemOpen
  \bibfield  {author} {\bibinfo {author} {\bibfnamefont {H.}~\bibnamefont {Shouval}},\ }\bibfield  {title} {\bibinfo {title} {Spike timing dependent plasticity: A consequence of more fundamental learning rules},\ }\bibfield  {journal} {\bibinfo  {journal} {Frontiers in Computational Neuroscience}\ }\href {https://doi.org/10.3389/fncom.2010.00019} {10.3389/fncom.2010.00019} (\bibinfo {year} {2010})\BibitemShut {NoStop}%
\bibitem [{\citenamefont {Song}\ \emph {et~al.}(2000)\citenamefont {Song}, \citenamefont {Miller},\ and\ \citenamefont {Abbott}}]{song_competitive_2000-1}%
  \BibitemOpen
  \bibfield  {author} {\bibinfo {author} {\bibfnamefont {S.}~\bibnamefont {Song}}, \bibinfo {author} {\bibfnamefont {K.~D.}\ \bibnamefont {Miller}},\ and\ \bibinfo {author} {\bibfnamefont {L.~F.}\ \bibnamefont {Abbott}},\ }\bibfield  {title} {\bibinfo {title} {Competitive {{Hebbian}} learning through spike-timing-dependent synaptic plasticity},\ }\href {https://doi.org/10.1038/78829} {\bibfield  {journal} {\bibinfo  {journal} {Nature Neuroscience}\ }\textbf {\bibinfo {volume} {3}},\ \bibinfo {pages} {919} (\bibinfo {year} {2000})}\BibitemShut {NoStop}%
\bibitem [{\citenamefont {Fiete}\ \emph {et~al.}(2010)\citenamefont {Fiete}, \citenamefont {Senn}, \citenamefont {Wang},\ and\ \citenamefont {Hahnloser}}]{Fiete2010}%
  \BibitemOpen
  \bibfield  {author} {\bibinfo {author} {\bibfnamefont {I.~R.}\ \bibnamefont {Fiete}}, \bibinfo {author} {\bibfnamefont {W.}~\bibnamefont {Senn}}, \bibinfo {author} {\bibfnamefont {C.~Z.}\ \bibnamefont {Wang}},\ and\ \bibinfo {author} {\bibfnamefont {R.~H.}\ \bibnamefont {Hahnloser}},\ }\bibfield  {title} {\bibinfo {title} {Spike-time-dependent plasticity and heterosynaptic competition organize networks to produce long scale-free sequences of neural activity},\ }\href {https://doi.org/10.1016/j.neuron.2010.02.003} {\bibfield  {journal} {\bibinfo  {journal} {Neuron}\ }\textbf {\bibinfo {volume} {65}},\ \bibinfo {pages} {563–576} (\bibinfo {year} {2010})}\BibitemShut {NoStop}%
\bibitem [{\citenamefont {Jambon-Puillet}\ \emph {et~al.}(2024)\citenamefont {Jambon-Puillet}, \citenamefont {Testa}, \citenamefont {Lorenz}, \citenamefont {Style}, \citenamefont {Rebane},\ and\ \citenamefont {Dufresne}}]{JambonPuillet2024}%
  \BibitemOpen
  \bibfield  {author} {\bibinfo {author} {\bibfnamefont {E.}~\bibnamefont {Jambon-Puillet}}, \bibinfo {author} {\bibfnamefont {A.}~\bibnamefont {Testa}}, \bibinfo {author} {\bibfnamefont {C.}~\bibnamefont {Lorenz}}, \bibinfo {author} {\bibfnamefont {R.~W.}\ \bibnamefont {Style}}, \bibinfo {author} {\bibfnamefont {A.~A.}\ \bibnamefont {Rebane}},\ and\ \bibinfo {author} {\bibfnamefont {E.~R.}\ \bibnamefont {Dufresne}},\ }\bibfield  {title} {\bibinfo {title} {Phase-separated droplets swim to their dissolution},\ }\bibfield  {journal} {\bibinfo  {journal} {Nature Communications}\ }\textbf {\bibinfo {volume} {15}},\ \href {https://doi.org/10.1038/s41467-024-47889-y} {10.1038/s41467-024-47889-y} (\bibinfo {year} {2024})\BibitemShut {NoStop}%
\bibitem [{\citenamefont {Fredriksson}\ \emph {et~al.}(2002)\citenamefont {Fredriksson}, \citenamefont {Gullberg}, \citenamefont {Jarvius}, \citenamefont {Olsson}, \citenamefont {Pietras}, \citenamefont {Gústafsdóttir}, \citenamefont {Östman},\ and\ \citenamefont {Landegren}}]{Fredriksson2002}%
  \BibitemOpen
  \bibfield  {author} {\bibinfo {author} {\bibfnamefont {S.}~\bibnamefont {Fredriksson}}, \bibinfo {author} {\bibfnamefont {M.}~\bibnamefont {Gullberg}}, \bibinfo {author} {\bibfnamefont {J.}~\bibnamefont {Jarvius}}, \bibinfo {author} {\bibfnamefont {C.}~\bibnamefont {Olsson}}, \bibinfo {author} {\bibfnamefont {K.}~\bibnamefont {Pietras}}, \bibinfo {author} {\bibfnamefont {S.~M.}\ \bibnamefont {Gústafsdóttir}}, \bibinfo {author} {\bibfnamefont {A.}~\bibnamefont {Östman}},\ and\ \bibinfo {author} {\bibfnamefont {U.}~\bibnamefont {Landegren}},\ }\bibfield  {title} {\bibinfo {title} {Protein detection using proximity-dependent dna ligation assays},\ }\href {https://doi.org/10.1038/nbt0502-473} {\bibfield  {journal} {\bibinfo  {journal} {Nature Biotechnology}\ }\textbf {\bibinfo {volume} {20}},\ \bibinfo {pages} {473–477} (\bibinfo {year} {2002})}\BibitemShut {NoStop}%
\bibitem [{\citenamefont {Schaus}\ \emph {et~al.}(2017)\citenamefont {Schaus}, \citenamefont {Woo}, \citenamefont {Xuan}, \citenamefont {Chen},\ and\ \citenamefont {Yin}}]{Schaus2017}%
  \BibitemOpen
  \bibfield  {author} {\bibinfo {author} {\bibfnamefont {T.~E.}\ \bibnamefont {Schaus}}, \bibinfo {author} {\bibfnamefont {S.}~\bibnamefont {Woo}}, \bibinfo {author} {\bibfnamefont {F.}~\bibnamefont {Xuan}}, \bibinfo {author} {\bibfnamefont {X.}~\bibnamefont {Chen}},\ and\ \bibinfo {author} {\bibfnamefont {P.}~\bibnamefont {Yin}},\ }\bibfield  {title} {\bibinfo {title} {A dna nanoscope via auto-cycling proximity recording},\ }\bibfield  {journal} {\bibinfo  {journal} {Nature Communications}\ }\textbf {\bibinfo {volume} {8}},\ \href {https://doi.org/10.1038/s41467-017-00542-3} {10.1038/s41467-017-00542-3} (\bibinfo {year} {2017})\BibitemShut {NoStop}%
\bibitem [{\citenamefont {Moerman}\ \emph {et~al.}(2023)\citenamefont {Moerman}, \citenamefont {Fang}, \citenamefont {Videbæk}, \citenamefont {Rogers},\ and\ \citenamefont {Schulman}}]{Moerman2023}%
  \BibitemOpen
  \bibfield  {author} {\bibinfo {author} {\bibfnamefont {P.~G.}\ \bibnamefont {Moerman}}, \bibinfo {author} {\bibfnamefont {H.}~\bibnamefont {Fang}}, \bibinfo {author} {\bibfnamefont {T.~E.}\ \bibnamefont {Videbæk}}, \bibinfo {author} {\bibfnamefont {W.~B.}\ \bibnamefont {Rogers}},\ and\ \bibinfo {author} {\bibfnamefont {R.}~\bibnamefont {Schulman}},\ }\bibfield  {title} {\bibinfo {title} {A simple method to alter the binding specificity of dna-coated colloids that crystallize},\ }\href {https://doi.org/10.1039/d3sm01105d} {\bibfield  {journal} {\bibinfo  {journal} {Soft Matter}\ }\textbf {\bibinfo {volume} {19}},\ \bibinfo {pages} {8779–8789} (\bibinfo {year} {2023})}\BibitemShut {NoStop}%
\bibitem [{\citenamefont {Weinstein}\ \emph {et~al.}(2019)\citenamefont {Weinstein}, \citenamefont {Regev},\ and\ \citenamefont {Zhang}}]{Weinstein2019}%
  \BibitemOpen
  \bibfield  {author} {\bibinfo {author} {\bibfnamefont {J.~A.}\ \bibnamefont {Weinstein}}, \bibinfo {author} {\bibfnamefont {A.}~\bibnamefont {Regev}},\ and\ \bibinfo {author} {\bibfnamefont {F.}~\bibnamefont {Zhang}},\ }\bibfield  {title} {\bibinfo {title} {Dna microscopy: Optics-free spatio-genetic imaging by a stand-alone chemical reaction},\ }\href {https://doi.org/10.1016/j.cell.2019.05.019} {\bibfield  {journal} {\bibinfo  {journal} {Cell}\ }\textbf {\bibinfo {volume} {178}},\ \bibinfo {pages} {229} (\bibinfo {year} {2019})}\BibitemShut {NoStop}%
\bibitem [{\citenamefont {Owen}\ \emph {et~al.}(2023)\citenamefont {Owen}, \citenamefont {Osmanović},\ and\ \citenamefont {Mirny}}]{Owen2023}%
  \BibitemOpen
  \bibfield  {author} {\bibinfo {author} {\bibfnamefont {J.~A.}\ \bibnamefont {Owen}}, \bibinfo {author} {\bibfnamefont {D.}~\bibnamefont {Osmanović}},\ and\ \bibinfo {author} {\bibfnamefont {L.}~\bibnamefont {Mirny}},\ }\bibfield  {title} {\bibinfo {title} {Design principles of 3d epigenetic memory systems},\ }\bibfield  {journal} {\bibinfo  {journal} {Science}\ }\textbf {\bibinfo {volume} {382}},\ \href {https://doi.org/10.1126/science.adg3053} {10.1126/science.adg3053} (\bibinfo {year} {2023})\BibitemShut {NoStop}%
\bibitem [{\citenamefont {Falk}\ \emph {et~al.}(2023{\natexlab{b}})\citenamefont {Falk}, \citenamefont {Strupp}, \citenamefont {Scellier},\ and\ \citenamefont {Murugan}}]{Falk2023}%
  \BibitemOpen
  \bibfield  {author} {\bibinfo {author} {\bibfnamefont {M.}~\bibnamefont {Falk}}, \bibinfo {author} {\bibfnamefont {A.}~\bibnamefont {Strupp}}, \bibinfo {author} {\bibfnamefont {B.}~\bibnamefont {Scellier}},\ and\ \bibinfo {author} {\bibfnamefont {A.}~\bibnamefont {Murugan}},\ }\href@noop {} {\bibinfo {title} {Contrastive learning through non-equilibrium memory}} (\bibinfo {year} {2023}{\natexlab{b}}),\ \Eprint {https://arxiv.org/abs/2312.17723} {arXiv:2312.17723} \BibitemShut {NoStop}%
\bibitem [{\citenamefont {Hopfield}(1982)}]{Hopfield1982}%
  \BibitemOpen
  \bibfield  {author} {\bibinfo {author} {\bibfnamefont {J.~J.}\ \bibnamefont {Hopfield}},\ }\bibfield  {title} {\bibinfo {title} {Neural networks and physical systems with emergent collective computational abilities.},\ }\href {https://doi.org/10.1073/pnas.79.8.2554} {\bibfield  {journal} {\bibinfo  {journal} {Proceedings of the National Academy of Sciences}\ }\textbf {\bibinfo {volume} {79}},\ \bibinfo {pages} {2554–2558} (\bibinfo {year} {1982})}\BibitemShut {NoStop}%
\bibitem [{\citenamefont {Ramsauer}\ \emph {et~al.}(2020)\citenamefont {Ramsauer}, \citenamefont {Schäfl}, \citenamefont {Lehner}, \citenamefont {Seidl}, \citenamefont {Widrich}, \citenamefont {Adler}, \citenamefont {Gruber}, \citenamefont {Holzleitner}, \citenamefont {Pavlović}, \citenamefont {Sandve}, \citenamefont {Greiff}, \citenamefont {Kreil}, \citenamefont {Kopp}, \citenamefont {Klambauer}, \citenamefont {Brandstetter},\ and\ \citenamefont {Hochreiter}}]{Ramsauer2020}%
  \BibitemOpen
  \bibfield  {author} {\bibinfo {author} {\bibfnamefont {H.}~\bibnamefont {Ramsauer}}, \bibinfo {author} {\bibfnamefont {B.}~\bibnamefont {Schäfl}}, \bibinfo {author} {\bibfnamefont {J.}~\bibnamefont {Lehner}}, \bibinfo {author} {\bibfnamefont {P.}~\bibnamefont {Seidl}}, \bibinfo {author} {\bibfnamefont {M.}~\bibnamefont {Widrich}}, \bibinfo {author} {\bibfnamefont {T.}~\bibnamefont {Adler}}, \bibinfo {author} {\bibfnamefont {L.}~\bibnamefont {Gruber}}, \bibinfo {author} {\bibfnamefont {M.}~\bibnamefont {Holzleitner}}, \bibinfo {author} {\bibfnamefont {M.}~\bibnamefont {Pavlović}}, \bibinfo {author} {\bibfnamefont {G.~K.}\ \bibnamefont {Sandve}}, \bibinfo {author} {\bibfnamefont {V.}~\bibnamefont {Greiff}}, \bibinfo {author} {\bibfnamefont {D.}~\bibnamefont {Kreil}}, \bibinfo {author} {\bibfnamefont {M.}~\bibnamefont {Kopp}}, \bibinfo {author} {\bibfnamefont {G.}~\bibnamefont {Klambauer}}, \bibinfo {author} {\bibfnamefont {J.}~\bibnamefont {Brandstetter}},\ and\ \bibinfo {author} {\bibfnamefont
  {S.}~\bibnamefont {Hochreiter}},\ }\href@noop {} {\bibinfo {title} {Hopfield networks is all you need}} (\bibinfo {year} {2020}),\ \Eprint {https://arxiv.org/abs/2008.02217} {arXiv:2008.02217} \BibitemShut {NoStop}%
\bibitem [{\citenamefont {Hopfield}(1984)}]{Hopfield1984}%
  \BibitemOpen
  \bibfield  {author} {\bibinfo {author} {\bibfnamefont {J.~J.}\ \bibnamefont {Hopfield}},\ }\bibfield  {title} {\bibinfo {title} {Neurons with graded response have collective computational properties like those of two-state neurons.},\ }\href {https://doi.org/10.1073/pnas.81.10.3088} {\bibfield  {journal} {\bibinfo  {journal} {Proceedings of the National Academy of Sciences}\ }\textbf {\bibinfo {volume} {81}},\ \bibinfo {pages} {3088–3092} (\bibinfo {year} {1984})}\BibitemShut {NoStop}%
\bibitem [{\citenamefont {Koiran}(1994)}]{Koiran1994}%
  \BibitemOpen
  \bibfield  {author} {\bibinfo {author} {\bibfnamefont {P.}~\bibnamefont {Koiran}},\ }\bibfield  {title} {\bibinfo {title} {Dynamics of discrete time, continuous state hopfield networks},\ }\href {https://doi.org/10.1162/neco.1994.6.3.459} {\bibfield  {journal} {\bibinfo  {journal} {Neural Computation}\ }\textbf {\bibinfo {volume} {6}},\ \bibinfo {pages} {459–468} (\bibinfo {year} {1994})}\BibitemShut {NoStop}%
\bibitem [{\citenamefont {Clark}\ and\ \citenamefont {Abbott}(2024)}]{Clark2024}%
  \BibitemOpen
  \bibfield  {author} {\bibinfo {author} {\bibfnamefont {D.~G.}\ \bibnamefont {Clark}}\ and\ \bibinfo {author} {\bibfnamefont {L.}~\bibnamefont {Abbott}},\ }\bibfield  {title} {\bibinfo {title} {Theory of coupled neuronal-synaptic dynamics},\ }\href {https://doi.org/10.1103/physrevx.14.021001} {\bibfield  {journal} {\bibinfo  {journal} {Physical Review X}\ }\textbf {\bibinfo {volume} {14}},\ \bibinfo {pages} {021001} (\bibinfo {year} {2024})}\BibitemShut {NoStop}%
\bibitem [{Note1()}]{Note1}%
  \BibitemOpen
  \bibinfo {note} {Note that when $\protect \bm {f}[\protect \bm {s},\protect \bm {J}]$ depends on the past trajectory of the system, then $\protect \bm {s}$ in the figure represents the full past trajectory of the system. In practice, when $\protect \bm {f}$ does not only depend on the values of $\protect \bm {s}(t')$ in the recent past, it approximated by the instantaneous values of $\protect \bm {s}$, $\protect \bm {\protect \dot s}$, $\protect \bm {\protect \ddot s}$, ... up to a certain order.}\BibitemShut {Stop}%
\bibitem [{\citenamefont {Fang}\ \emph {et~al.}(2019)\citenamefont {Fang}, \citenamefont {Kruse}, \citenamefont {Lu},\ and\ \citenamefont {Wang}}]{Fang2019}%
  \BibitemOpen
  \bibfield  {author} {\bibinfo {author} {\bibfnamefont {X.}~\bibnamefont {Fang}}, \bibinfo {author} {\bibfnamefont {K.}~\bibnamefont {Kruse}}, \bibinfo {author} {\bibfnamefont {T.}~\bibnamefont {Lu}},\ and\ \bibinfo {author} {\bibfnamefont {J.}~\bibnamefont {Wang}},\ }\bibfield  {title} {\bibinfo {title} {Nonequilibrium physics in biology},\ }\href {https://doi.org/10.1103/revmodphys.91.045004} {\bibfield  {journal} {\bibinfo  {journal} {Reviews of Modern Physics}\ }\textbf {\bibinfo {volume} {91}},\ \bibinfo {pages} {045004} (\bibinfo {year} {2019})}\BibitemShut {NoStop}%
\bibitem [{{\relax DLMF}()}]{NIST_DLMF}%
  \BibitemOpen
  {\relax DLMF},\ \href {https://dlmf.nist.gov/} {\bibinfo {title} {{\it NIST Digital Library of Mathematical Functions}}},\ \bibinfo {howpublished} {\url{https://dlmf.nist.gov/}, Release 1.1.10 of 2023-06-15} (\bibinfo {year} {2023}),\ \bibinfo {note} {f.~W.~J. Olver, A.~B. {Olde Daalhuis}, D.~W. Lozier, B.~I. Schneider, R.~F. Boisvert, C.~W. Clark, B.~R. Miller, B.~V. Saunders, H.~S. Cohl, and M.~A. McClain, eds.}\BibitemShut {Stop}%
\end{thebibliography}%
\end{document}